\documentclass[10pt,final,journal,twocolumn]{IEEEtran}
\ifCLASSINFOpdf

\else
\usepackage{algorithm}
\usepackage{booktabs}

\fi
\usepackage{cite}
\usepackage[cmex10]{amsmath}
\usepackage{amsthm}
\usepackage{stfloats}
\usepackage{multicol,multienum}
\usepackage{multirow}
\usepackage{mdframed}
\usepackage{amssymb}
\usepackage{url}
\usepackage{amsmath}
\usepackage{xcolor}
\usepackage[dvips]{graphicx}
\usepackage{subfigure}
\usepackage{caption}
\usepackage{amsmath}
\usepackage{amsfonts,amssymb}
\usepackage{bbm}
\usepackage[T1]{fontenc}
\usepackage{algorithm}
\usepackage{algpseudocode}
\usepackage{ulem}
\usepackage{balance}
\usepackage[linesnumbered, ruled]{}
\makeatletter
\renewcommand{\maketag@@@}[1]{\hbox{\m@th\normalsize\normalfont#1}}%
\makeatother

\begin{document}
\hyphenation{op-tical net-works semi-conduc-tor}
\title{\vspace{-0.35em} Outage-Aware Sum Rate Maximization in Movable Antennas-Enabled Systems}
\author{Guojie Hu, Qingqing Wu, Ming-Min Zhao, Wen Chen, Zhenyu Xiao, Kui Xu and Jiangbo Si\vspace{-1.54em}
\thanks{
%
Guojie Hu is with the Department of Electronic Engineering, Shanghai Jiao Tong University, Shanghai 200240, China, and also with the College of Communication Engineering, Rocket Force University of Engineering, Xi'an 710025, China (lgdxhgj@sina.com). Qingqing Wu and Wen Chen are with the Department of Electronic Engineering, Shanghai Jiao Tong University, Shanghai 200240, China. Ming-Min Zhao is with the College of Information Science and Electronic Engineering, Zhejiang University, Hangzhou 310058, China. Zhenyu Xiao is with the School of Electronic and Information Engineering and State Key Laboratory of CNS/ATM,
 Beihang University, Beijing 100191, China. Kui Xu is with the College of Communications Engineering, Army Engineering University of PLA, Nanjing 210007, China. Jiangbo Si is with the Integrated Service Networks Lab of Xidian University, Xi'an 710100, China.
}
}
\IEEEpeerreviewmaketitle
\maketitle
\begin{abstract}
In this paper, we investigate the movable antennas (MAs)-enabled multiple-input-single-output (MISO) systems, where the base station (BS) equipped with multiple MAs serves multiple single-antenna user. The delay-sensitive scenario is considered, where users refrain from periodically sending training signals to the BS for channel estimations to avoid additional latency. As a result, the BS relies solely on the statistical channel state information (CSI) to transmit data with a fixed rate. Under this setup, we aim to maximize the \textit{outage-aware} sum rate of all users, by jointly optimizing antenna positions and the transmit beamforming at the BS, while satisfying the given target outage probability requirement at each user. The problem is highly non-convex, primarily because the exact cumulative distribution function (CDF) of the received signal-to-interference-plus-noise ratio (SINR) of each user is difficult to derive. To simplify analysis and without comprising performance, we adopt the statistical CSI based zero-forcing beamforming design. We then introduce one important lemma to derive the tight mean and variance of the SINR. Leveraging these results, we further exploit the Laguerre series approximation to successfully derive the closed-form and tight CDF of the SINR. Subsequently, the outage-aware sum rate expression is presented but still includes complex structure with respect to antenna positions. Facing this challenge, the projected gradient ascent (PGA) method is developed to iteratively update antenna positions until convergence. Numerical results demonstrate the effectiveness of our proposed schemes compared to conventional fixed-position antenna (FPA) and other competitive benchmarks.
\end{abstract}

\IEEEpeerreviewmaketitle
\vspace{-10pt}
\section{Introduction}
The multiuser downlink communication is a fundamental scenario in wireless systems, where the base station (BS) can simultaneously transmit distinct signals (such as video and voice) to multiple users over the same time-frequency resources. The multiple-input-multiple-output (MIMO) technology, serving as a key enabler in this context, has played a pivotal role in boosting spectral efficiency by several orders of magnitude compared to single-antenna systems \cite{An_Overview, Shift_MIMO}.

 Conventional MIMO systems typically exploit fixed-position antennas (FPAs) with a constant inter-antenna spacing of half-wavelength \cite{Massivea, Massivet}. This rigid architecture, however, will cause a fundamental limitation in reconfiguring wireless channel conditions. Therefore, in scenarios where users are distributed in close proximity, i.e., the correlations between the corresponding downlink channels are strong, the FPAs-enabled BS may not be able to unleash the full potential of its transmit beamforming \cite{Guojie1}. A natural approach to mitigate this issue is to increase the number of BS antennas, as seen in massive MIMO \cite{Xiqi_Gao, Massive2} or even extremely large-scale antenna array (ELAA) systems \cite{DLL1, DLL2}, which enhances channel orthogonality, diversity and spatial multiplexing capabilities. However, massive MIMO or ELAA presents a significant drawback: The need for numerous antennas and radio frequency (RF) chains leads to substantially high energy consumption. This raises a critical question: Is there a technology that can achieve higher spectral efficiency while consuming less energy?

 \begin{figure}
\centering
\includegraphics[width=7.5cm]{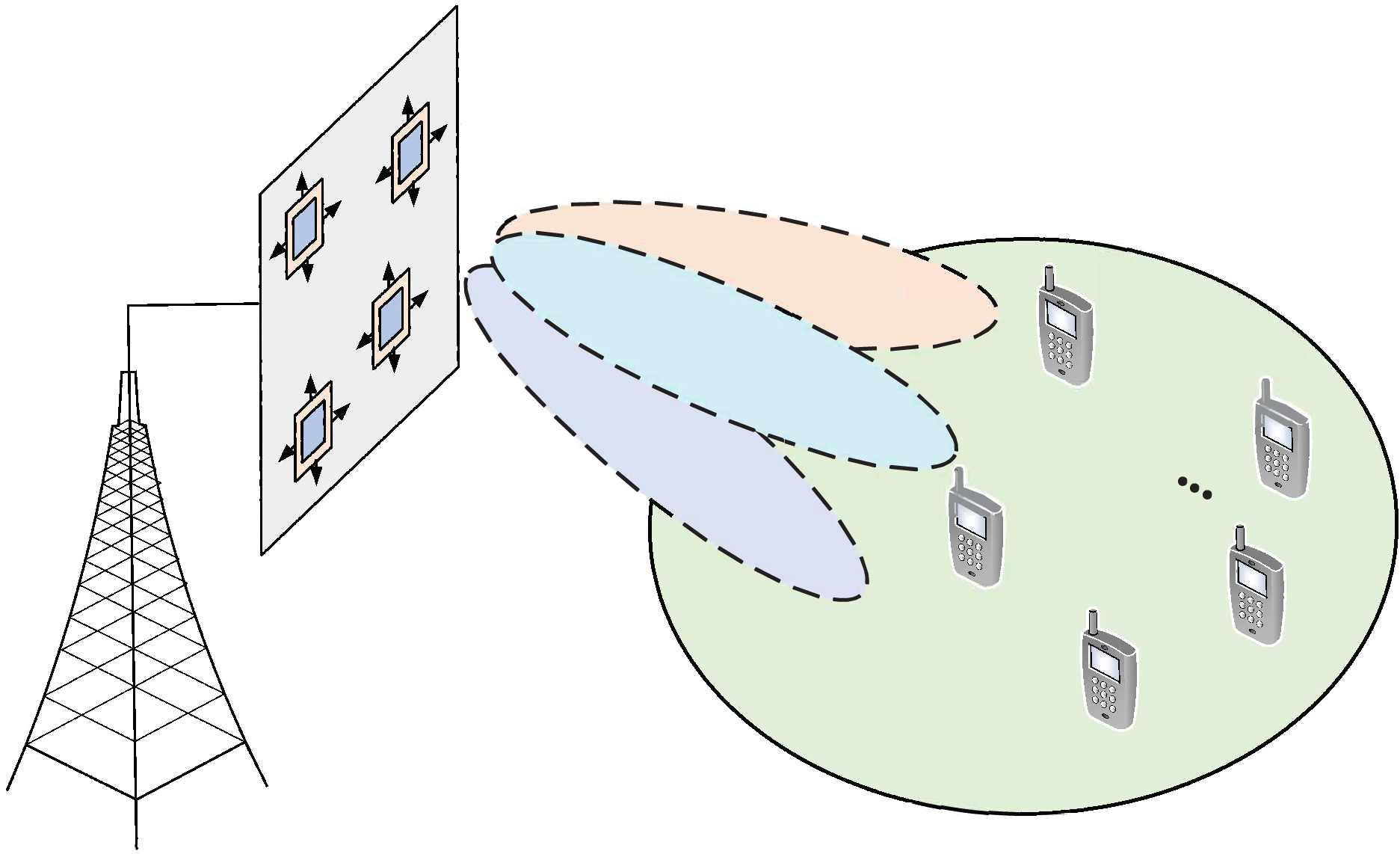}
\captionsetup{font=small}
\caption{Illustration of the considered system model.} \label{fig:Fig1}
\vspace{-15pt}
\end{figure}


Recently, movable antennas (MAs) have emerged as a promising solution \cite{ZLP_survey, Qingqing_arxiv}. Unlike traditional FPAs, each MA is connected via a flexible cable and can be dynamically repositioned by a mechanical driver over a continuous region spanning several to tens of wavelengths. This mobility brings an additional spatial degrees of freedom (DoF), and provides advanced channel reconfiguration capability, allowing the system to create favorable prerequisites for the beamforming designs and then achieve significant performance gains without increasing the number of antennas or RF chains \cite{Z_WCM, ZLP_MODEL, MWYY_TWC, gao2025integrating, Wanghaohao}. Given its advantages, significant research efforts have been devoted to applying the MA technology in multiuser communication scenarios. For instance, i) for the multiuser uplink, \cite{ZLP23} considered that there are multiple single-MA users transmitting signals to the FPAs-enabled BS, employing the projected gradient descent (PGD) method to optimize antenna positions for total transmit power minimization. Concurrently, \cite{GUOJIE_CL} proposed a low-complexity PGD method for handling the same problem. In a similar setup, \cite{XZY1} developed a two-loop iterative algorithm for solving antenna positions, the receive combining matrix of the BS and the transmit power of each user for maximizing the minimum rate among users. Different from \cite{ZLP23, GUOJIE_CL, XZY1}, \cite{GUOJIE_TWOTIMESCALE} proposed a two-timescale transmission protocol, where antenna positions are first optimized based on the statistical channel state information (CSI), while the receive combining matrix is subsequently solved with the instantaneous CSI. A more complex setting was considered in \cite{XUHAO_TCOM}, where MAs-enabled users simultaneously communicate with the FPAs-enabled BS, and the capacity is maximized by jointly optimizing transmit covariance matrices and antenna positions. A new six-dimensional movable antenna (6DMA) architecture was proposed in \cite{XIAODAN1, XIAODAN2} to provide full DoF for improving the sum rate to the maximum extent; ii) for the multiuser downlink, \cite{WUYIFEI_TCOM} developed a branch and bound method to jointly optimize the transmit beamforming and antenna positions at the BS for minimizing the total BS power consumption while guaranteeing a minimum signal-to-interference-plus-noise ratio (SINR) for each user. \cite{IOT} further integrated MAs into flexible cylindrical arrays for improving coverage, energy efficiency and flexibility. With the enlargement of the Rayleigh distance due to antenna movements, \cite{MA_NEAR1, MA_NEAR2} emphasized the necessity of adopting near-field spherical-wave models in MA systems. While \cite{ZHENZY} and \cite{CHENXT} further introduced the two-timescale transmission protocol into multiple-input-single-output (MISO) and MIMO systems, respectively.

The aforementioned studies all focused on delay-tolerant communication scenarios. In such scenarios, the system is not sensitive to delays. This allows all users to transmit pilot signals to the BS in each transmission slot. Subsequently, the BS can estimate the instantaneous CSI by detecting these signals and then perform the system design based on the acquired instantaneous CSI. Nevertheless, the above process about ``Feedback-Estimate-Transmit'' will be not suitable for delay-sensitive communication scenarios such as video streaming, online gaming, and other popular 5G services \cite{Delay_sensitive}, as such scenarios require stringent latency constraints. Instead, in delay-sensitive communication scenarios, it is more advisable for the BS to continuously transmit data relying on the statistical CSI, which varies slowly and can be evaluated once and for all before long-term transmissions. This paradigm shift implies that for MAs-enabled multiuser systems, there will be no feedback from user to the BS for channel estimations, and the design of both antenna positions and the transmit beamforming should be based solely on the statistical CSI, which, as will be elaborated, poses significant challenges to the optimization process.

Different from the aforementioned studies, we in this paper investigate MAs-enabled MISO systems under delay-sensitive communication scenarios. The systems comprise one BS equipped with multiple MAs serving multiple single-antenna users over Rician fading channels. Since the BS has access only to the statistical CSI, it should adopt the fixed rate transmission mode, which in turn leads to random outage events during transmission \cite{Outage1, Outage2}. Considering this fact, our objective is to maximize the \textit{outage-aware} sum rate by jointly optimizing antenna positions and the transmit beamforming at the BS, subject to the given outage probability requirements at users. The formulated problem is highly non-convex, mainly due to two aspects: i) Antenna positions and the transmit beamforming are intricately coupled in the objective; ii) The cumulative distribution function (CDF) of the received SINR is difficult to derive, thereby obstructing the path to obtain the closed-form expression for the outage-aware sum rate. Motivated by these challenges, we in this paper develop a comprehensive solution framework, as summarized below.
\begin{itemize}
\item[$\bullet$] First, given antenna positions at the BS, we adopt the statistical CSI based zero-forcing (ZF) beamforming design for simplifying the SINR expression without compromising performance. Unfortunately, the resulting SINR remains a ratio of two correlated and positive random variables, which complicates the derivation of its exact CDF. Facing this difficulty, we introduce one important lemma that leverages both first- and second-order Taylor expansion theory, based on which the tight mean and variance of the SINR are successfully obtained.

 \item[$\bullet$] Second, leveraging the tight approximations of the SINR's mean and variance, we obtain its CDF in closed form using the Laguerre series approximation. This allows us to subsequently obtain a closed-form expression for the outage-aware sum rate through mathematical transformations. However, the resulting expression still involves the inverse of the incomplete gamma function. To address this, we propose a novel linear approximation for this inverse function, thereby reformulating the original problem into a simpler form with a tractable structure, which is only related to antenna positions at the BS.

 \item[$\bullet$] Third, although more tractable, the simplified problem remains highly non-convex because the antenna position variables are intricately embedded within complex-valued vectors and matrices. Considering that the gradient method does not require the objective function to be convex or concave and can efficiently handle multivariable optimization in a low-complexity manner, we develop a projected gradient ascent (PGA) framework. This framework iteratively optimizes antenna positions until obtaining a high-quality stationary solution.

 \item[$\bullet$] Finally, we conduct numerical simulations and present several competitive benchmarks to demonstrate the effectiveness and robustness of our proposed scheme across a wide range of system parameters.
\end{itemize}

The rest of this paper is organized as follows. Section II introduces the system model and then presents the problem formulation. Section III aims to derive the approximated CDF of the SINR expression. Section VI provides the process of obtaining the outage-aware rate expression. Section V focuses the detailed optimization algorithm. Simulation results are provided in Section VI and finally conclusions are given in Section VII.


 \newcounter{mytempeqncnt}
\section{System Model and Problem Formulation}
\subsection{System Model}
As shown in Fig. 1, we consider the MAs-enabled downlink multiuser communications, in which the BS equipped with $N$ MAs serves $M$ single-antenna users, thereby forming the MISO architecture. Without loss of generality, we denote ${\bf{t}} \buildrel \Delta \over = \left\{ {{{\bf{t}}_1},{{\bf{t}}_2},...,{{\bf{t}}_N}} \right\} \in {{\mathbb{R}}^{2 \times N}}$ as the positions of $N$ MAs at the BS, where ${{\bf{t}}_n} = {\left[ {{x_n},{y_n}} \right]^T}$ represents the $n$-th MA's position relative to the reference point ${[0,0]^T}$, with ${{\bf{t}}_n} \in {{\cal A}_n}$ and ${{\cal A}_n} \buildrel \Delta \over = \left\{ {{x_n} \in \left\{ {x_n^{\min },x_n^{\max }} \right\},{y_n} \in \left\{ {y_n^{\min },y_n^{\min }} \right\}} \right\}$ is the moving region of the $n$-th MA. In particular, in order to avoid the coupling effect, there is no overlapping area between ${{\cal A}_n}$ and ${{\cal A}_j}$ for any $n \ne j$. In addition, it is assumed that the movement region of MAs is significantly smaller than the signal propagation distance. Therefore, the far-field assumption holds between the BS and the users, which indicates that adjusting MA positions only alters the phase of the multi-path components, while the angles of departure (AoD), angles of arrival (AoA) and channel gain amplitudes remain unchanged \cite{ZLP23, ZHENZY, CHENXT}.

The equivalent baseband channel from the BS to the $m$-th user is denoted by ${{\bf{h}}_m}({\bf{t}}) \in {{\mathbb{C}}^{1 \times N}}$, which is characterized with the general Rician fading model, i.e.,
\begin{equation}
\begin{split}{}
{{\bf{h}}_m}({\bf{t}}) = \sqrt {\frac{{{K_m}{\beta _m}}}{{{K_m} + 1}}} {\overline {\bf{h}} _m}({\bf{t}}) + \sqrt {\frac{{{\beta _m}}}{{{K_m} + 1}}} {\widetilde {\bf{h}}_m}({\bf{t}}),
\end{split}
\end{equation}
where ${K_m} > 0$ is the Rician K-factor, ${{\beta _m}}$ is the large-scale fading coefficient, and ${\overline {\bf{h}} _m}({\bf{t}}) \in {{\mathbb{C}}^{1 \times N}}$ and ${\widetilde {\bf{h}}_m}({\bf{t}}) \in {{\mathbb{C}}^{1 \times N}}$ are the line-of-sight (LoS) component and random non-LoS (NLoS) component, respectively. Specifically, ${\overline {\bf{h}} _m}({\bf{t}})$ can be expressed as
\begin{equation}
\begin{split}{}
{\overline {\bf{h}} _m}({\bf{t}}) = \left[ {{e^{j\frac{{2\pi }}{\lambda }{\bf{t}}_1^T{{\bf{a}}_m}}},...,{e^{j\frac{{2\pi }}{\lambda }{\bf{t}}_N^T{{\bf{a}}_m}}}} \right],
\end{split}
\end{equation}
where $\lambda $ is the carrier wavelength, ${{\bf{a}}_m} = {\left[ {\cos {\theta _m}\sin {\phi _m},\sin {\theta _m}} \right]^T}$, and ${\theta _m} \in \left[ { - \frac{\pi }{2},\frac{\pi }{2}} \right]$ and ${\phi _m} \in \left[ { - \frac{\pi }{2},\frac{\pi }{2}} \right]$ denote the elevation and azimuth AoDs for the $m$-th user, respectively. In addition, entries of ${\widetilde {\bf{h}}_m}({\bf{t}})$ are independent and identically distributed (i.i.d.) circularly symmetric complex Gaussian random variables with zero mean and unit variance.

Let $P_m$ and ${{\bf{w}}_m} \in {{\mathbb{C}}^{N \times 1}}$ respectively denote the transmit power and the beamforming vector at the BS for serving the $m$-th user, with $\left\| {{{\bf{w}}_m}} \right\| = 1$, $\forall m \in {\cal M} \buildrel \Delta \over = \left\{ {1,...,M} \right\}$. Hence, the received signal at the $m$-th user is given by
\begin{equation}
\begin{split}{}
{y_m} =& \sqrt {{P_m}} {{\bf{h}}_m}({\bf{t}}){{\bf{w}}_m}{s_m}\\
 &+ \sum\nolimits_{j = 1,j \ne m}^M {\sqrt {{P_j}} {{\bf{h}}_m}({\bf{t}}){{\bf{w}}_j}{s_j} + {n_m}},
\end{split}
\end{equation}
where ${{s_m}}$ is the information-bearing signal for the $m$-th user with ${\mathbb{E}}\left[ {{{\left| {{s_m}} \right|}^2}} \right] = 1$, $\forall m \in {\cal M}$, and ${n_m} \sim {\cal C}{\cal N}(0,{\sigma ^2})$ represents the additive white Gaussian noise (AWGN) at the $m$-th user. Based on (3), the received SINR of the $m$-th user is derived as
\begin{equation}
\begin{split}{}
{\gamma _m}({\bf{t}},{\bf{W}}) = \frac{{{P_m}{{\left| {{{\bf{h}}_m}({\bf{t}}){{\bf{w}}_m}} \right|}^2}}}{{\sum\nolimits_{j = 1,j \ne m}^M {{P_j}{{\left| {{{\bf{h}}_m}({\bf{t}}){{\bf{w}}_j}} \right|}^2} + {\sigma ^2}} }},
\end{split}
\end{equation}
where ${\bf{W}} \buildrel \Delta \over = \left[ {{{\bf{w}}_1},...,{{\bf{w}}_M}} \right] \in {{\mathbb{C}}^{N \times M}}$.

In this paper, we focus on delay-sensitive communication applications such as video data transmissions. In such applications, users do not periodically send training signals to the BS for channel estimations, as this process would introduce additional latency. Instead, the BS acquires only the statistical CSI of all downlink channels via established methods such as global positioning systems, and subsequently relies on this information to perform the system design. Specifically, in the absence of instantaneous CSI, the BS is unable to dynamically adjust its transmission rate on a per-slot basis and should instead employ a fixed rate for each user. Under this setup, there would exist the outage event in the signal transmission process. Hence, for a given outage probability requirement $\delta $, the transmission rate (denoted by $R_m$) for the $m$-th user should be the solution of the following equation
\begin{equation}
\begin{split}{}
\Pr \left\{ {{{\log }_2}\left( {1 + {\gamma _m}({\bf{t}},{\bf{W}})} \right) < {R_m}} \right\} = \delta.
\end{split}
\end{equation}

We aim to maximize the outage-aware sum rate $\sum\nolimits_{m = 1}^M {{R_m}} $, via the joint design of antenna positions ${\bf{t}}$ and the beamforming matrix ${\bf{W}}$ at the BS. Therefore, the optimization problem can be formulated as
 \begin{align}
&({\rm{P1}}):{\rm{  }}\mathop {\max }\limits_{{\bf{t}},{\bf{W}},\left\{ {{R_m}} \right\}_{m = 1}^M} \ \sum\nolimits_{m = 1}^M {{R_m}}  \tag{${\rm{6a}}$}\\
{\rm{              }}&{\rm{s.t.}} \ {{\bf{t}}_n} \in {{\cal A}_n},\forall n \in {\cal N},\tag{${\rm{6b}}$}\\
 &\ \ \ \ \left\| {{{\bf{w}}_m}} \right\| = 1,\forall m \in {\cal M},\tag{${\rm{6c}}$}\\
 &\ \ \ \ \ {\rm{Pr}} \left\{ {{{\log }_2}\left( {1 + {\gamma _m}({\bf{t}},{\bf{W}})} \right) < {R_m}} \right\} = \delta, \forall m \in {\cal M}, \tag{${\rm{6d}}$}
\end{align}
where ${\cal N} \buildrel \Delta \over = \left\{ {1,...,N} \right\}$.

\textbf{Remark 1:} Note that both antenna positions ${\bf{t}}$ and the beamforming matrix ${\bf{W}}$ at the BS are designed based solely on the statistical CSI. As a result, it becomes challenging to derive the CDF of the SINR ${\gamma _m}({\bf{t}},{\bf{W}})$, $\forall m \in {\cal M}$, as well as the outage probability expression shown in (6d). Moreover, ${\bf{t}}$ and ${\bf{W}}$ are highly coupled with each other in ${\gamma _m}({\bf{t}},{\bf{W}})$, $\forall m \in {\cal M}$. Consequently, problem (P1) becomes highly non-convex and difficult to solve.

\section{The Approximated CDF of ${\gamma _m}({\bf{t}},{{\bf{W}}})$}
Via the above analysis, to efficiently solve (P1), the first task is to derive the CDF of ${{\gamma _m}({\bf{t}},{\bf{W}})}$. To proceed, we in this paper adopt the ZF-based transmit beamforming for facilitating analysis and without compromising performance. Specifically, given antenna positions ${\bf{t}}$, the statistical CSI can be characterized by $\overline {\bf{H}} ({\bf{t}}) \buildrel \Delta \over = {\left[ {{{\overline {\bf{h}} }_1}({\bf{t}});...;{{\overline {\bf{h}} }_M}({\bf{t}})} \right]^H} \in {{\mathbb{C}}^{N \times M}}$. Relying on the statistical CSI, the ZF-based beamforming for the $m$-th user can expressed as ${\bf{w}}_m^{{\rm{ZF}}}({\bf{t}}) = {{\bf{w}}_m}({\bf{t}})/\left\| {{{\bf{w}}_m}({\bf{t}})} \right\|$ \cite{GUOJIE_TWOTIMESCALE}, with
\begin{equation}
\setcounter{equation}{7}
\begin{split}{}
{{\bf{w}}_m}({\bf{t}}) = \left( {{{\bf{I}}_N} - {{\overline {\bf{H}} }_m}({\bf{t}}){{\left( {\overline {\bf{H}} _m^H({\bf{t}}){{\overline {\bf{H}} }_m}({\bf{t}})} \right)}^{ - 1}}\overline {\bf{H}} _m^H({\bf{t}})} \right)\overline {\bf{h}} _m^H({\bf{t}}),
\end{split}
\end{equation}
where ${\overline {\bf{H}} _m}({\bf{t}}) = {[{\overline {\bf{h}} _1}({\bf{t}});...,{\overline {\bf{h}} _{m - 1}}({\bf{t}});{\overline {\bf{h}} _{m + 1}}({\bf{t}});...;{\overline {\bf{h}} _M}({\bf{t}})]^H} \in {{\mathbb{C}}^{N \times (M - 1)}}$. Substituting ${{\bf{W}}^{{\rm{ZF}}}}({\bf{t}}) \buildrel \Delta \over = \left[ {{\bf{w}}_1^{{\rm{ZF}}}({\bf{t}}),...,{\bf{w}}_M^{{\rm{ZF}}}({\bf{t}})} \right]$ into (4), ${\gamma _m}({\bf{t}},{{\bf{W}}^{{\rm{ZF}}}}({\bf{t}}))$ can be expanded as
\begin{equation}
\begin{split}{}
&{\gamma _m}({\bf{t}},{{\bf{W}}^{{\rm{ZF}}}}({\bf{t}}))\\
 =& \frac{{{P_m}{{\left| {\sqrt {\frac{{{K_m}{\beta _m}}}{{{K_m} + 1}}} {{\overline {\bf{h}} }_m}({\bf{t}}){\bf{w}}_m^{{\rm{ZF}}}({\bf{t}}) + \sqrt {\frac{{{\beta _m}}}{{{K_m} + 1}}} {{\widetilde {\bf{h}}}_m}({\bf{t}}){\bf{w}}_m^{{\rm{ZF}}}({\bf{t}})} \right|}^2}}}{{\sum\nolimits_{j = 1,j \ne m}^M {{P_j}{{\left| {\sqrt {\frac{{{\beta _m}}}{{{K_m} + 1}}} {{\widetilde {\bf{h}}}_m}({\bf{t}}){\bf{w}}_j^{{\rm{ZF}}}({\bf{t}})} \right|}^2} + {\sigma ^2}} }},
\end{split}
\end{equation}
where the denominator in (8) has such form mainly because ${\overline {\bf{h}} _m}({\bf{t}}){\bf{w}}_j^{{\rm{ZF}}}({\bf{t}}) = 0$, $\forall j \in {\cal M}$ and $j \ne m$.

Even the ZF-based beamforming is adopted, it is still difficult to derive the exact CDF of ${\gamma _m}({\bf{t}},{{\bf{W}}^{{\rm{ZF}}}}({\bf{t}}))$ due to its complex structure. Therefore, in the next, we aim to find the tight approximation for the CDF expression. To proceed, we first provide one important conclusion in the follows.

\textbf{Lemma 1:} Let $Z = \frac{X}{Y}$, where $X > 0$ ($Y > 0$) is a random variable with the mean and variance being ${\mathbb{E}}[X]$ (${\mathbb{E}}[Y]$) and ${\mathbb{V}}[X]$ (${\mathbb{V}}[Y]$), respectively, and $X$ and $Y$ are correlated with the covariance being ${\rm{Cov}}(X,Y)$. Then, the mean and variance of the random variable $Z$ can be tightly approximated as
\begin{equation}
\begin{split}{}
{\mathbb{E}}[Z] \approx &\frac{{{\mathbb{E}}\left[ X \right]}}{{{\mathbb{E}}\left[ Y \right]}} + \frac{{{\mathbb{E}}\left[ X \right]}}{{{{\left( {{\mathbb{E}}\left[ Y \right]} \right)}^3}}}{\mathbb{V}}\left[ Y \right] - \frac{{{\rm{Cov}}(X,Y)}}{{{{\left( {{\mathbb{E}}\left[ Y \right]} \right)}^2}}},\\
{\mathbb{V}}[Z] \approx &\frac{{{\mathbb{V}}\left[ X \right]}}{{{{\left( {{\mathbb{E}}\left[ Y \right]} \right)}^2}}} + \frac{{{{\left( {{\mathbb{E}}\left[ X \right]} \right)}^2}}}{{{{\left( {{\mathbb{E}}\left[ Y \right]} \right)}^4}}}{\mathbb{V}}\left[ Y \right] - \frac{{2{\mathbb{E}}\left[ X \right]{\rm{Cov}}(X,Y)}}{{{{\left( {{\mathbb{E}}\left[ Y \right]} \right)}^3}}}.
\end{split}
\end{equation}

\begin{proof}
Let us first perform the first-order Taylor expansion on the variable $Z = \frac{X}{Y}$ around $\left( {{\mathbb{E}}[X],{\mathbb{E}}[Y]} \right)$, which can be expressed as
\begin{equation}
\begin{split}{}
Z \approx& \frac{{{\mathbb{E}}\left[ X \right]}}{{{\mathbb{E}}\left[ Y \right]}} + \frac{{\partial Z}}{{\partial X}}{|_{X = {\mathbb{E}}\left[ X \right],Y = {\mathbb{E}}\left[ Y \right]}}(X - {\mathbb{E}}\left[ X \right])\\
{\rm{     }}& + \frac{{\partial Z}}{{\partial Y}}{|_{X = {\mathbb{E}}\left[ X \right],Y = {\mathbb{E}}\left[ Y \right]}}(Y - {\mathbb{E}}\left[ Y \right])\\
 =& \frac{{{\mathbb{E}}\left[ X \right]}}{{{\mathbb{E}}\left[ Y \right]}} + \frac{1}{{{\mathbb{E}}\left[ Y \right]}}(X - {\mathbb{E}}\left[ X \right])\\
 &- \frac{{{\mathbb{E}}\left[ X \right]}}{{{{\left( {{\mathbb{E}}\left[ Y \right]} \right)}^2}}}(Y - {\mathbb{E}}\left[ Y \right]) \buildrel \Delta \over = {Z_{{\rm{first - order}}}},
\end{split}
\end{equation}
based on which we can obtain
\begin{equation}
\begin{split}{}
{\mathbb{V}}[Z] \approx &{\mathbb{V}}\left[ {\frac{X}{{{\mathbb{E}}\left[ Y \right]}}} \right] + {\mathbb{V}}\left[ {\frac{{{\mathbb{E}}\left[ X \right]}}{{{{\left( {{\mathbb{E}}\left[ Y \right]} \right)}^2}}}Y} \right] \\
&+ 2{\rm{Cov}}\left( {\frac{1}{{{\mathbb{E}}\left[ Y \right]}}X, - \frac{{{\mathbb{E}}\left[ X \right]}}{{{{\left( {{\mathbb{E}}\left[ Y \right]} \right)}^2}}}Y} \right),
\end{split}
\end{equation}
by further substituting the following results into (11),
\begin{equation} \nonumber
\begin{split}{}
{\mathbb{V}}\left[ {\frac{X}{{{\mathbb{E}}\left[ Y \right]}}} \right] =& \frac{{{\mathbb{V}}\left[ X \right]}}{{{{\left( {{\mathbb{E}}\left[ Y \right]} \right)}^2}}},\\
{\mathbb{V}}\left[ {\frac{{{\mathbb{E}}\left[ X \right]}}{{{{\left( {{\mathbb{E}}\left[ Y \right]} \right)}^2}}}Y} \right] =& \frac{{{{\left( {{\mathbb{E}}\left[ X \right]} \right)}^2}}}{{{{\left( {{\mathbb{E}}\left[ Y \right]} \right)}^4}}}{\mathbb{V}}\left[ Y \right],\\
{\rm{Cov}}\left( {\frac{1}{{{\mathbb{E}}\left[ Y \right]}}X, - \frac{{{\mathbb{E}}\left[ X \right]}}{{{{\left( {{\mathbb{E}}\left[ {\rm{Y}} \right]} \right)}^2}}}Y} \right) =&  - \frac{{{\mathbb{E}}\left[ X \right]}}{{{{\left( {{\mathbb{E}}\left[ Y \right]} \right)}^3}}}{\rm{Cov}}\left( {X,Y} \right),
\end{split}
\end{equation}
we can obtain the tight approximation about the variance of $Z$ shown in (9).

Generally, the first-order Taylor expansion method is sufficient for deriving the tight variance of the variable $Z$. However, such method is not sufficient to derive the tight mean of the variable $Z$, which will be illustrated later. Therefore, the second-order Taylor method is further employed to obtain a more tight mean. Specifically, the second-order Taylor expansion on the variable $Z = \frac{X}{Y}$ around $\left( {{\mathbb{E}}[X],{\mathbb{E}}[Y]} \right)$ is
\begin{equation}
\begin{split}{}
Z \approx& {Z_{{\rm{first - order}}}} + \frac{1}{2}\left( {\frac{{{\partial ^2}Z}}{{\partial {X^2}}}{|_{X = {\mathbb{E}}\left[ X \right],Y = {\mathbb{E}}\left[ Y \right]}}{{(X - {\mathbb{E}}\left[ X \right])}^2}} \right.\\
 &+ 2\frac{{{\partial ^2}Z}}{{\partial X\partial Y}}{|_{X = {\mathbb{E}}\left[ X \right],Y = {\mathbb{E}}\left[ Y \right]}}(X - {\mathbb{E}}\left[ X \right])(Y - {\mathbb{E}}\left[ Y \right])  \\
&+\left. {\frac{{{\partial ^2}Z}}{{\partial {Y^2}}}{|_{X = {\mathbb{E}}\left[ X \right],Y = {\mathbb{E}}\left[ Y \right]}}{{(Y - {\mathbb{E}}\left[ Y \right])}^2}} \right),
\end{split}
\end{equation}
by further substituting the following results into (12),
 \begin{equation} \nonumber
\begin{split}{}
{\mathbb{E}}\left[ {{Z_{{\rm{first - order}}}}} \right] =& \frac{{{\mathbb{E}}\left[ X \right]}}{{{\mathbb{E}}\left[ Y \right]}},\\
\frac{{{\partial ^2}Z}}{{\partial {X^2}}}{|_{X = {\mathbb{E}}\left[ X \right],Y = {\mathbb{E}}\left[ Y \right]}} =& 0,\\
\frac{{{\partial ^2}Z}}{{\partial X\partial Y}}{|_{X = {\mathbb{E}}\left[ X \right],Y = {\mathbb{E}}\left[ Y \right]}} =& \frac{{ - 1}}{{{{\left( {{\mathbb{E}}\left[ Y \right]} \right)}^2}}},
\end{split}
\end{equation}
 \begin{equation} \nonumber
\begin{split}{}
{\mathbb{E}}\left[ {(X - {\mathbb{E}}\left[ X \right])(Y - {\mathbb{E}}\left[ Y \right])} \right] =& {\rm{Cov}}\left( {X,Y} \right),\\
\frac{{{\partial ^2}Z}}{{\partial {Y^2}}}{|_{X = {\mathbb{E}}\left[ X \right],Y = {\mathbb{E}}\left[ Y \right]}} = &\frac{{2{\mathbb{E}}\left[ X \right]}}{{{{\left( {{\mathbb{E}}\left[ Y \right]} \right)}^3}}},\\
{\mathbb{E}}\left[ {{{(Y - {\mathbb{E}}\left[ Y \right])}^2}} \right] =& {\mathbb{V}}\left[ Y \right],
\end{split}
\end{equation}
we can finally obtain the tight approximation about the mean of $Z$ shown in (9). This completes the proof.
\end{proof}

Based on Lemma 1, we now define that
\begin{equation}
\begin{split}{}
{\gamma _m}({\bf{t}},{{\bf{W}}^{{\rm{ZF}}}}({\bf{t}})) \buildrel \Delta \over = {Z_m} = {P_m}\frac{{{X_m}}}{{{Y_m}}},
\end{split}
\end{equation}
with
 \begin{equation} \nonumber
\begin{split}{}
{X_m} =& {\left| {\sqrt {\frac{{{K_m}{\beta _m}}}{{{K_m} + 1}}} {{\overline {\bf{h}} }_m}({\bf{t}}){\bf{w}}_m^{{\rm{ZF}}}({\bf{t}}) + \sqrt {\frac{{{\beta _m}}}{{{K_m} + 1}}} {{\widetilde {\bf{h}}}_m}({\bf{t}}){\bf{w}}_m^{{\rm{ZF}}}({\bf{t}})} \right|^2},\\
{Y_m} =& \sum\nolimits_{j = 1,j \ne m}^M {{P_j}{{\left| {\sqrt {\frac{{{\beta _m}}}{{{K_m} + 1}}} {{\widetilde {\bf{h}}}_m}({\bf{t}}){\bf{w}}_j^{{\rm{ZF}}}({\bf{t}})} \right|}^2} + {\sigma ^2}}.
\end{split}
\end{equation}

In the next, we aim to derive the mean, variance and the covariance of $X_m$ and $Y_m$, respectively, in preparation for deriving the CDF of $Z_m$.

\subsection{The expressions of ${\mathbb{E}}\left[ {{X_m}} \right]$ and ${\mathbb{V}}\left[ {{X_m}} \right]$}
For the random variable $X_m$, since ${{{\overline {\bf{h}} }_m}({\bf{t}}){\bf{w}}_m^{{\rm{ZF}}}({\bf{t}})}$ is a deterministic component and ${{{\widetilde {\bf{h}}}_m}({\bf{t}}){\bf{w}}_m^{{\rm{ZF}}}({\bf{t}})}$ is a circularly-symmetric complex Gaussian random component (due to that ${{\bf{w}}_m^{{\rm{ZF}}}({\bf{t}})}$ is independent of ${{{\widetilde {\bf{h}}}_m}({\bf{t}})}$) with zero mean and unit variance, it can be concluded that the mean (denoted by ${\mu _m}$) and variance (denoted by $\sigma _m^2$) of the complex variable ${\sqrt {\frac{{{K_m}{\beta _m}}}{{{K_m} + 1}}} {{\overline {\bf{h}} }_m}({\bf{t}}){\bf{w}}_m^{{\rm{ZF}}}({\bf{t}}) + \sqrt {\frac{{{\beta _m}}}{{{K_m} + 1}}} {{\widetilde {\bf{h}}}_m}({\bf{t}}){\bf{w}}_m^{{\rm{ZF}}}({\bf{t}})}$ are
  \begin{equation} \nonumber
\begin{split}{}
{\mu _m} =& \sqrt {\frac{{{K_m}{\beta _m}}}{{{K_m} + 1}}} {\overline {\bf{h}} _m}({\bf{t}}){\bf{w}}_m^{{\rm{ZF}}}({\bf{t}}),\\
\sigma _m^2 =& \frac{{{\beta _m}}}{{{K_m} + 1}}.
\end{split}
\end{equation}
 Therefore, the mean and variance of $X_m$ can be derived as
\begin{equation}
\begin{split}{}
{\mathbb{E}}\left[ {{X_m}} \right] =& {\left| {{\mu _m}} \right|^2} + \sigma _m^2\\
 =& \frac{{{K_m}{\beta _m}}}{{{K_m} + 1}}{\left| {{{\overline {\bf{h}} }_m}({\bf{t}}){\bf{w}}_m^{{\rm{ZF}}}({\bf{t}})} \right|^2} + \frac{{{\beta _m}}}{{{K_m} + 1}},\\
{\mathbb{V}}\left[ {{X_m}} \right] =& \sigma _m^4 + 2\sigma _m^2{\left| {{\mu _m}} \right|^2}\\
 =& \frac{{\beta _m^2}}{{{{\left( {{K_m} + 1} \right)}^2}}}\left( {1 + 2{K_m}{{\left| {{{\overline {\bf{h}} }_m}({\bf{t}}){\bf{w}}_m^{{\rm{ZF}}}({\bf{t}})} \right|}^2}} \right).
\end{split}
\end{equation}

\subsection{The expressions of ${\mathbb{E}}\left[ {{Y_m}} \right]$ and ${\mathbb{V}}\left[ {{Y_m}} \right]$}
For the random variable $Y_m$, note that it can be equivalently expressed as
\begin{equation}
\begin{split}{}
{Y_m} =& \sum\nolimits_{j = 1,j \ne m}^M {{P_j}{{\left| {\sqrt {\frac{{{\beta _m}}}{{{K_m} + 1}}} {{\widetilde {\bf{h}}}_m}({\bf{t}}){\bf{w}}_j^{{\rm{ZF}}}({\bf{t}})} \right|}^2} + {\sigma ^2}} \\
 =& \frac{{{\beta _m}}}{{{K_m} + 1}}{\widetilde {\bf{h}}_m}({\bf{t}})\left( {\underbrace {\sum\nolimits_{j = 1,j \ne m}^M {{P_j}{\bf{w}}_j^{{\rm{ZF}}}({\bf{t}}){{\left( {{\bf{w}}_j^{{\rm{ZF}}}({\bf{t}})} \right)}^H}} }_{{{\bf{\Psi }}_m}({\bf{t}})}} \right)\\
 &\times \widetilde {\bf{h}}_m^H({\bf{t}}) + {\sigma ^2}.
\end{split}
\end{equation}
where ${{{\bf{\Psi }}_m}({\bf{t}})}$ is a Hermite matrix with ${{\bf{\Psi }}_m}({\bf{t}}) = {\bf{\Psi }}_m^H({\bf{t}})$.

Then, since
\begin{equation}
\begin{split}{}
&{\mathbb{E}}\left[ {{{\widetilde {\bf{h}}}_m}({\bf{t}}){{\bf{\Psi }}_m}({\bf{t}})\widetilde {\bf{h}}_m^H({\bf{t}})} \right]\\
 =& {\rm{tr}}\left( {{{\bf{\Psi }}_m}({\bf{t}}){\mathbb{E}}\left[ {\widetilde {\bf{h}}_m^H({\bf{t}}){{\widetilde {\bf{h}}}_m}({\bf{t}})} \right]} \right)\\
 =& {\rm{tr}}\left( {{{\bf{\Psi }}_m}({\bf{t}}){{\bf{I}}_N}} \right) = {\rm{tr}}\left( {{{\bf{\Psi }}_m}({\bf{t}})} \right)\\
 =& \sum\nolimits_{j = 1,j \ne m}^M {{P_j}} {\rm{tr}}\left( {{\bf{w}}_j^{{\rm{ZF}}}({\bf{t}}){{\left( {{\bf{w}}_j^{{\rm{ZF}}}({\bf{t}})} \right)}^H}} \right)\\
 =& \sum\nolimits_{j = 1,j \ne m}^M {{P_j}},
\end{split}
\end{equation}
we can obtain the mean of $Y_m$ as
\begin{equation}
\begin{split}{}
{\mathbb{E}}\left[ {{Y_m}} \right] = \frac{{{\beta _m}}}{{{K_m} + 1}}\sum\nolimits_{j = 1,j \ne m}^M {{P_j}}  + {\sigma ^2}.
\end{split}
\end{equation}

\setcounter{equation}{\value{mytempeqncnt}}
  \begin{figure*}[b!]
  \hrulefill
\setcounter{equation}{25}
\begin{equation}  \footnotesize
\begin{split}{}
{\mathbb{E}}\left[ {{X_m}{Y_m}} \right] =& {\sigma ^2}{\mathbb{E}}\left[ {{X_m}} \right] + \sum\nolimits_{j = 1,j \ne m}^M {{P_j}{\mathbb{E}}\left[ {\underbrace {{{\left| {\underbrace {\sqrt {\frac{{{K_m}\beta _m^2}}{{{{\left( {{K_m} + 1} \right)}^2}}}} {{\overline {\bf{h}} }_m}({\bf{t}}){\bf{w}}_m^{{\rm{ZF}}}({\bf{t}}){{\widetilde {\bf{h}}}_m}({\bf{t}}){\bf{w}}_j^{{\rm{ZF}}}({\bf{t}})}_{{\varepsilon _{mj,1}}({\bf{t}})} + \underbrace {\frac{{{\beta _m}}}{{{K_m} + 1}}{{\widetilde {\bf{h}}}_m}({\bf{t}}){\bf{w}}_m^{{\rm{ZF}}}({\bf{t}}){{\widetilde {\bf{h}}}_m}({\bf{t}}){\bf{w}}_j^{{\rm{ZF}}}({\bf{t}})}_{{\varepsilon _{mj,2}}({\bf{t}})}} \right|}^2}}_{{\varepsilon _{mj}}({\bf{t}})}} \right]}.
\end{split}
\end{equation}
\end{figure*}

  \begin{figure*}[b!]
  \hrulefill
\setcounter{equation}{27}
\begin{equation}
\begin{split}{}
&{\rm{Cov}}\left( {{X_m},{Y_m}} \right) = \sum\nolimits_{j = 1,j \ne m}^M {{P_j}} \left( {\frac{{{K_m}\beta _m^2}}{{{{\left( {{K_m} + 1} \right)}^2}}}{{\left| {{{\overline {\bf{h}} }_m}({\bf{t}}){\bf{w}}_m^{{\rm{ZF}}}({\bf{t}})} \right|}^2} + \frac{{\beta _m^2}}{{{{\left( {{K_m} + 1} \right)}^2}}}\left( {1 + {{\left| {{{\left( {{\bf{w}}_j^{{\rm{ZF}}}({\bf{t}})} \right)}^H}{\bf{w}}_m^{{\rm{ZF}}}({\bf{t}})} \right|}^2}} \right)} \right)\\
 &+ {\sigma ^2}{\mathbb{E}}\left[ {{X_m}} \right] - {\mathbb{E}}\left[ {{X_m}} \right]\left( {\sum\nolimits_{j = 1,j \ne m}^M {{P_j}\frac{{{\beta _m}}}{{{K_m} + 1}} + {\sigma ^2}} } \right) = \sum\nolimits_{j = 1,j \ne m}^M {{P_j}\frac{{\beta _m^2}}{{{{\left( {{K_m} + 1} \right)}^2}}}{{\left| {{{\left( {{\bf{w}}_j^{{\rm{ZF}}}({\bf{t}})} \right)}^H}{\bf{w}}_m^{{\rm{ZF}}}({\bf{t}})} \right|}^2}}.
\end{split}
\end{equation}
\end{figure*}

On the other hand, to derive the variance of $Y_m$, we now expand ${\left( {{{\widetilde {\bf{h}}}_m}({\bf{t}}){{\bf{\Psi }}_m}({\bf{t}})\widetilde {\bf{h}}_m^H({\bf{t}})} \right)^2}$ as
\begin{equation}
\setcounter{equation}{18}
\begin{split}{}
&{\left( {{{\widetilde {\bf{h}}}_m}({\bf{t}}){{\bf{\Psi }}_m}({\bf{t}})\widetilde {\bf{h}}_m^H({\bf{t}})} \right)^2}\\
 =& \left( {\sum\nolimits_{i = 1}^N {\sum\nolimits_{j = 1}^N {{{\left[ {{{\bf{\Psi }}_m}({\bf{t}})} \right]}_{ij}}{{\left[ {{{\widetilde {\bf{h}}}_m}({\bf{t}})} \right]}_i}{{\left[ {\widetilde {\bf{h}}_m^H({\bf{t}})} \right]}_j}} } } \right)\\
 &\times \left( {\sum\nolimits_{k = 1}^N {\sum\nolimits_{l = 1}^N {{{\left[ {{{\bf{\Psi }}_m}({\bf{t}})} \right]}_{kl}}{{\left[ {{{\widetilde {\bf{h}}}_m}({\bf{t}})} \right]}_k}{{\left[ {\widetilde {\bf{h}}_m^H({\bf{t}})} \right]}_l}} } } \right),
\end{split}
\end{equation}
based on which we can derive that
\begin{equation}
\begin{split}{}
&{\mathbb{E}}\left[ {{{\left( {{{\widetilde {\bf{h}}}_m}({\bf{t}}){{\bf{\Psi }}_m}({\bf{t}})\widetilde {\bf{h}}_m^H({\bf{t}})} \right)}^2}} \right]\\
 =& \sum\nolimits_{i = 1}^N {\sum\nolimits_{j = 1}^N {\sum\nolimits_{k = 1}^N {\sum\nolimits_{l = 1}^N {{{\left[ {{{\bf{\Psi }}_m}({\bf{t}})} \right]}_{ij}}{{\left[ {{{\bf{\Psi }}_m}({\bf{t}})} \right]}_{kl}}} } } } \\
 &\times {\mathbb{E}}\left[ {{{\left[ {{{\widetilde {\bf{h}}}_m}({\bf{t}})} \right]}_i}{{\left[ {\widetilde {\bf{h}}_m^H({\bf{t}})} \right]}_j}{{\left[ {{{\widetilde {\bf{h}}}_m}({\bf{t}})} \right]}_k}{{\left[ {\widetilde {\bf{h}}_m^H({\bf{t}})} \right]}_l}} \right].
\end{split}
\end{equation}
Further, since
\begin{equation}
\begin{split}{}
&{\mathbb{E}}\left[ {{{\left[ {{{\widetilde {\bf{h}}}_m}({\bf{t}})} \right]}_i}{{\left[ {\widetilde {\bf{h}}_m^H({\bf{t}})} \right]}_j}{{\left[ {{{\widetilde {\bf{h}}}_m}({\bf{t}})} \right]}_k}{{\left[ {\widetilde {\bf{h}}_m^H({\bf{t}})} \right]}_l}} \right]\\
 =& {\mathbb{E}}\left[ {{{\left[ {{{\widetilde {\bf{h}}}_m}({\bf{t}})} \right]}_i}{{\left[ {\widetilde {\bf{h}}_m^H({\bf{t}})} \right]}_j}} \right]{\mathbb{E}}\left[ {{{\left[ {{{\widetilde {\bf{h}}}_m}({\bf{t}})} \right]}_k}{{\left[ {\widetilde {\bf{h}}_m^H({\bf{t}})} \right]}_l}} \right]\\
 &+ {\mathbb{E}}\left[ {{{\left[ {{{\widetilde {\bf{h}}}_m}({\bf{t}})} \right]}_i}{{\left[ {\widetilde {\bf{h}}_m^H({\bf{t}})} \right]}_l}} \right]{\mathbb{E}}\left[ {{{\left[ {{{\widetilde {\bf{h}}}_m}({\bf{t}})} \right]}_k}{{\left[ {\widetilde {\bf{h}}_m^H({\bf{t}})} \right]}_j}} \right]\\
 &+ {\mathbb{E}}\left[ {{{\left[ {{{\widetilde {\bf{h}}}_m}({\bf{t}})} \right]}_i}{{\left[ {{{\widetilde {\bf{h}}}_m}({\bf{t}})} \right]}_k}} \right]{\mathbb{E}}\left[ {{{\left[ {\widetilde {\bf{h}}_m^H({\bf{t}})} \right]}_j}{{\left[ {\widetilde {\bf{h}}_m^H({\bf{t}})} \right]}_l}} \right],
\end{split}
\end{equation}
and
\begin{equation}  \nonumber
\begin{split}{}
{\mathbb{E}}\left[ {{{\left[ {{{\widetilde {\bf{h}}}_m}({\bf{t}})} \right]}_i}{{\left[ {\widetilde {\bf{h}}_m^H({\bf{t}})} \right]}_j}} \right] =& {\delta _{ij}},\\
{\mathbb{E}}\left[ {{{\left[ {{{\widetilde {\bf{h}}}_m}({\bf{t}})} \right]}_i}{{\left[ {{{\widetilde {\bf{h}}}_m}({\bf{t}})} \right]}_j}} \right] =& 0,\forall i,j\\
{\mathbb{E}}\left[ {{{\left[ {\widetilde {\bf{h}}_m^H({\bf{t}})} \right]}_i}{{\left[ {\widetilde {\bf{h}}_m^H({\bf{t}})} \right]}_j}} \right] =& 0,\forall i,j,
\end{split}
\end{equation}
with
\begin{equation} \nonumber
\begin{split}{}
{\delta _{ij}} = \left\{ {\begin{array}{*{20}{c}}
1&{i = j}\\
0&{i \ne j}
\end{array}} \right.,
\end{split}
\end{equation}
we can derive that
\begin{equation}
\begin{split}{}
&{\mathbb{E}}\left[ {{{\left( {{{\widetilde {\bf{h}}}_m}({\bf{t}}){{\bf{\Psi }}_m}({\bf{t}})\widetilde {\bf{h}}_m^H({\bf{t}})} \right)}^2}} \right]\\
 =& \sum\nolimits_{i = 1}^N {\sum\nolimits_{j = 1}^N {\sum\nolimits_{k = 1}^N {\sum\nolimits_{l = 1}^N {{{\left[ {{{\bf{\Psi }}_m}({\bf{t}})} \right]}_{ij}}{{\left[ {{{\bf{\Psi }}_m}({\bf{t}})} \right]}_{kl}}} } } } {\delta _{ij}}{\delta _{kl}}\\
 &+ \sum\nolimits_{i = 1}^N {\sum\nolimits_{j = 1}^N {\sum\nolimits_{k = 1}^N {\sum\nolimits_{l = 1}^N {{{\left[ {{{\bf{\Psi }}_m}({\bf{t}})} \right]}_{ij}}{{\left[ {{{\bf{\Psi }}_m}({\bf{t}})} \right]}_{kl}}} } } } {\delta _{il}}{\delta _{kj}},
\end{split}
\end{equation}
where the first term in (21) can be equivalently expressed as
\begin{equation} \nonumber
\begin{split}{}
&\sum\nolimits_{i = 1}^N {\sum\nolimits_{j = 1}^N {\sum\nolimits_{k = 1}^N {\sum\nolimits_{l = 1}^N {{{\left[ {{{\bf{\Psi }}_m}({\bf{t}})} \right]}_{ij}}{{\left[ {{{\bf{\Psi }}_m}({\bf{t}})} \right]}_{kl}}} } } } {\delta _{ij}}{\delta _{kl}}\\
 =& \sum\nolimits_{i = 1}^N {\sum\nolimits_{k = 1}^N {{{\left[ {{{\bf{\Psi }}_m}({\bf{t}})} \right]}_{ii}}{{\left[ {{{\bf{\Psi }}_m}({\bf{t}})} \right]}_{kk}}} } \\
 =& \left( {\sum\nolimits_{i = 1}^N {{{\left[ {{{\bf{\Psi }}_m}({\bf{t}})} \right]}_{ii}}} } \right)\left( {\sum\nolimits_{k = 1}^N {{{\left[ {{{\bf{\Psi }}_m}({\bf{t}})} \right]}_{kk}}} } \right)\\
 =& {\left( {{\rm{tr}}\left( {{{\bf{\Psi }}_m}({\bf{t}})} \right)} \right)^2},
\end{split}
\end{equation}
and the second term in (21) can be equivalently expressed as
\begin{equation} \nonumber
\begin{split}{}
&\sum\nolimits_{i = 1}^N {\sum\nolimits_{j = 1}^N {\sum\nolimits_{k = 1}^N {\sum\nolimits_{l = 1}^N {{{\left[ {{{\bf{\Psi }}_m}({\bf{t}})} \right]}_{ij}}{{\left[ {{{\bf{\Psi }}_m}({\bf{t}})} \right]}_{kl}}} } } } {\delta _{il}}{\delta _{kj}}\\
 =& \sum\nolimits_{i = 1}^N {\sum\nolimits_{j = 1}^N {{{\left[ {{{\bf{\Psi }}_m}({\bf{t}})} \right]}_{ij}}{{\left[ {{{\bf{\Psi }}_m}({\bf{t}})} \right]}_{ji}}} } \\
 =& \sum\nolimits_{i = 1}^N {\sum\nolimits_{j = 1}^N {{{\left[ {{{\bf{\Psi }}_m}({\bf{t}})} \right]}_{ij}}{{\left( {{{\left[ {{{\bf{\Psi }}_m}({\bf{t}})} \right]}_{ij}}} \right)}^H}} } \\
 =& {\rm{tr}}\left( {{{\left( {{{\bf{\Psi }}_m}({\bf{t}})} \right)}^2}} \right).
\end{split}
\end{equation}

According to the above analysis, we have
\begin{equation}
\begin{split}{}
&{\mathbb{E}}\left[ {{{\left( {{{\widetilde {\bf{h}}}_m}({\bf{t}}){{\bf{\Psi }}_m}({\bf{t}})\widetilde {\bf{h}}_m^H({\bf{t}})} \right)}^2}} \right]\\
 =& {\left( {{\rm{tr}}\left( {{{\bf{\Psi }}_m}({\bf{t}})} \right)} \right)^2} + {\rm{tr}}\left( {{{\left( {{{\bf{\Psi }}_m}({\bf{t}})} \right)}^2}} \right).
\end{split}
\end{equation}

Therefore, according to (16) and (22), we can derive the variance of ${{{\widetilde {\bf{h}}}_m}({\bf{t}}){{\bf{\Psi }}_m}({\bf{t}})\widetilde {\bf{h}}_m^H({\bf{t}})}$ as
\begin{equation}
\begin{split}{}
&{\mathbb{V}}\left[ {{{\widetilde {\bf{h}}}_m}({\bf{t}}){{\bf{\Psi }}_m}({\bf{t}})\widetilde {\bf{h}}_m^H({\bf{t}})} \right]\\
 =& {\mathbb{E}}\left[ {{{\left( {{{\widetilde {\bf{h}}}_m}({\bf{t}}){{\bf{\Psi }}_m}({\bf{t}})\widetilde {\bf{h}}_m^H({\bf{t}})} \right)}^2}} \right]\\
 &- {\left( {{\mathbb{E}}\left[ {{{\widetilde {\bf{h}}}_m}({\bf{t}}){{\bf{\Psi }}_m}({\bf{t}})\widetilde {\bf{h}}_m^H{{({\bf{t}})}^2}} \right]} \right)^2}\\
 =& {\rm{tr}}\left( {{{\left( {{{\bf{\Psi }}_m}({\bf{t}})} \right)}^2}} \right),
\end{split}
\end{equation}
leading to the variance of $Y_m$ as
\begin{equation}
\begin{split}{}
{\mathbb{V}}\left[ {{Y_m}} \right] = {\left( {\frac{{{\beta _m}}}{{{K_m} + 1}}} \right)^2}{\rm{tr}}\left( {{{\left( {{{\bf{\Psi }}_m}({\bf{t}})} \right)}^2}} \right).
\end{split}
\end{equation}

\subsection{The expressions of ${\rm{Cov}}\left( {{X_m},{Y_m}} \right)$}
Note that ${\rm{Cov}}\left( {{X_m},{Y_m}} \right)$ can be equivalently expressed as
\begin{equation}
\begin{split}{}
{\rm{Cov}}\left( {{X_m},{Y_m}} \right) = {\mathbb{E}}\left[ {{X_m}{Y_m}} \right] - {\mathbb{E}}\left[ {{X_m}} \right]{\mathbb{E}}\left[ {{Y_m}} \right],
\end{split}
\end{equation}
where ${\mathbb{E}}\left[ {{X_m}{Y_m}} \right]$ can be expanded as in (26), with
\begin{equation}
\setcounter{equation}{27}
\begin{split}{}
{\mathbb{E}}\left[ {{\varepsilon _{mj}}({\bf{t}})} \right] =& {\mathbb{E}}\left[ {{{\left| {{\varepsilon _{mj,1}}({\bf{t}})} \right|}^2}} \right] + {\mathbb{E}}\left[ {{{\left| {{\varepsilon _{mj,2}}({\bf{t}})} \right|}^2}} \right]\\
 &+ 2{\mathop{\rm Re}\nolimits} \left( {{\mathbb{E}}\left[ {{\varepsilon _{mj,1}}({\bf{t}})\varepsilon _{mj,2}^H({\bf{t}})} \right]} \right),
\end{split}
\end{equation}
and
\begin{equation} \nonumber
\begin{split}{}
&{\mathbb{E}}\left[ {{{\left| {{\varepsilon _{mj,1}}({\bf{t}})} \right|}^2}} \right] = \frac{{{K_m}\beta _m^2}}{{{{\left( {{K_m} + 1} \right)}^2}}}{\left| {{{\overline {\bf{h}} }_m}({\bf{t}}){\bf{w}}_m^{{\rm{ZF}}}({\bf{t}})} \right|^2},\\
&{\mathbb{E}}\left[ {{{\left| {{\varepsilon _{mj,2}}({\bf{t}})} \right|}^2}} \right] = \frac{{\beta _m^2}}{{{{\left( {{K_m} + 1} \right)}^2}}}\\
 &\times \left( {{\mathbb{E}}\left[ {{{\left| {{{\widetilde {\bf{h}}}_m}({\bf{t}}){\bf{w}}_m^{{\rm{ZF}}}({\bf{t}})} \right|}^2}} \right]{\mathbb{E}}\left[ {{{\left| {{{\widetilde {\bf{h}}}_m}({\bf{t}}){\bf{w}}_j^{{\rm{ZF}}}({\bf{t}})} \right|}^2}} \right]} \right.\\
&\left. { + {{\left| {{\mathbb{E}}\left[ {{{\widetilde {\bf{h}}}_m}({\bf{t}}){\bf{w}}_m^{{\rm{ZF}}}({\bf{t}}){{\left( {{{\widetilde {\bf{h}}}_m}({\bf{t}}){\bf{w}}_j^{{\rm{ZF}}}({\bf{t}})} \right)}^H}} \right]} \right|}^2}} \right)\\
 &= \frac{{\beta _m^2}}{{{{\left( {{K_m} + 1} \right)}^2}}}\left( {1 + {{\left| {{{\left( {{\bf{w}}_j^{{\rm{ZF}}}({\bf{t}})} \right)}^H}{\bf{w}}_m^{{\rm{ZF}}}({\bf{t}})} \right|}^2}} \right).
\end{split}
\end{equation}

In addition, note that
\begin{equation} \nonumber
\begin{split}{}
&{\rm{Re}}\left( {{\mathbb{E}}\left[ {{\varepsilon _{mj,1}}({\bf{t}})\varepsilon _{mj,2}^H({\bf{t}})} \right]} \right)\\
 =& \frac{{\sqrt {{K_m}} \beta _m^2}}{{{{\left( {{K_m} + 1} \right)}^2}}}{\mathop{\rm Re}\nolimits} \left( {{{\overline {\bf{h}} }_m}({\bf{t}}){\bf{w}}_m^{{\rm{ZF}}}({\bf{t}})} \right.\\
 &\times {\mathbb{E}}\left[ {{{\widetilde {\bf{h}}}_m}({\bf{t}}){\bf{w}}_j^{{\rm{ZF}}}({\bf{t}}){{\left( {{{\widetilde {\bf{h}}}_m}({\bf{t}}){\bf{w}}_m^{{\rm{ZF}}}({\bf{t}}){{\widetilde {\bf{h}}}_m}({\bf{t}}){\bf{w}}_j^{{\rm{ZF}}}({\bf{t}})} \right)}^H}} \right],
\end{split}
\end{equation}
with
\begin{equation} \nonumber
\begin{split}{}
&{\mathbb{E}}\left[ {{{\widetilde {\bf{h}}}_m}({\bf{t}}){\bf{w}}_j^{{\rm{ZF}}}({\bf{t}}){{\left( {{{\widetilde {\bf{h}}}_m}({\bf{t}}){\bf{w}}_m^{{\rm{ZF}}}({\bf{t}}){{\widetilde {\bf{h}}}_m}({\bf{t}}){\bf{w}}_j^{{\rm{ZF}}}({\bf{t}})} \right)}^H}} \right]\\
 =& \sum\nolimits_{i = 1}^N {\sum\nolimits_{o = 1}^N {\sum\nolimits_{k = 1}^N {\left( {{\bf{w}}_m^{{\rm{ZF}}}({\bf{t}})} \right)} } } _i^H{\left( {{\bf{w}}_j^{{\rm{ZF}}}({\bf{t}})} \right)_o}\left( {{\bf{w}}_j^{{\rm{ZF}}}({\bf{t}})} \right)_k^H\\
& \times {\mathbb{E}}\left[ {\left( {{{\widetilde {\bf{h}}}_m}({\bf{t}})} \right)_i^H{{\left( {{{\widetilde {\bf{h}}}_m}({\bf{t}})} \right)}_o}\left( {{{\widetilde {\bf{h}}}_m}({\bf{t}})} \right)_k^H} \right],
\end{split}
\end{equation}
where ${\mathbb{E}}\left[ {\left( {{{\widetilde {\bf{h}}}_m}({\bf{t}})} \right)_i^H{{\left( {{{\widetilde {\bf{h}}}_m}({\bf{t}})} \right)}_o}\left( {{{\widetilde {\bf{h}}}_m}({\bf{t}})} \right)_k^H} \right] = 0$, $\forall i,o,k$. Based on this fact, we can easily derive that
\begin{equation} \nonumber
\begin{split}{}
{\mathop{\rm Re}\nolimits} \left( {{\mathbb{E}}\left[ {{\varepsilon _{mj,1}}({\bf{t}})\varepsilon _{mj,2}^H({\bf{t}})} \right]} \right) = 0.
\end{split}
\end{equation}

\begin{table*}[t]
\caption{The values of ${\kappa (\delta )}$ and $\rho (\delta )$ with respect to $\delta  \in [0.1:0.01:0.2]$.}
\renewcommand\arraystretch{1.35}
\normalsize
\centering
\begin{tabular}[l]{|c|c|c|c|c|c|c|c|c|c|c|c|c|c|}
 \hline
 $\delta$ &$0.1$&$0.11$&$0.12$&$0.13$&$0.14$&$0.15$&$0.16$&$0.17$&$0.18$&$0.19$&$0.2$\\
 \hline
 $\kappa (\delta )$ &$0.7655$&$0.7752$&$0.7842$&$0.7928$&$0.801$&$0.8088$&$0.8163$&$0.8235$&$0.8304$&$0.8371$&$0.8437$\\
 \hline
${\rho (\delta )}$&$-1.188$&$-1.167$&$-1.145$&$-1.124$&$-1.103$&$-1.082$&$-1.061$&$-1.041$&$-1.02$&$-0.9993$&$-0.9787$\\
 \hline
 \end{tabular}
{
 \label{tab:tb1}
}
 \vspace{-10pt}
\end{table*}

  \begin{figure*}[b!]
  \hrulefill
\setcounter{equation}{35}
\begin{equation}
\begin{split}{}
{R_m} \approx {\log _2}\left( \begin{array}{l}
1 + \kappa (\delta ){P_m}\left( {\frac{{{\mathbb{E}}\left[ {{X_m}} \right]}}{{{\mathbb{E}}\left[ {{Y_m}} \right]}} + \frac{{{\mathbb{E}}\left[ {{X_m}} \right]}}{{{{\left( {{\mathbb{E}}\left[ {{Y_m}} \right]} \right)}^3}}}{\mathbb{V}}\left[ {{Y_m}} \right] - \frac{{{\rm{Cov}}({X_m},{Y_m})}}{{{{\left( {{\mathbb{E}}\left[ {{Y_m}} \right]} \right)}^2}}}} \right)\\
 + \rho (\delta )\frac{{{P_m}}}{{{\mathbb{E}}\left[ {{Y_m}} \right]}}\left( {\underbrace {\frac{{{\mathbb{V}}\left[ {{X_m}} \right] + \frac{{{{\left( {{\mathbb{E}}\left[ {{X_m}} \right]} \right)}^2}}}{{{{\left( {{\mathbb{E}}\left[ {{Y_m}} \right]} \right)}^2}}}{\mathbb{V}}\left[ {{Y_m}} \right] - 2\frac{{{\mathbb{E}}\left[ {{X_m}} \right]{\rm{Cov}}({X_m},{Y_m})}}{{{\mathbb{E}}\left[ {{Y_m}} \right]}}}}{{{\mathbb{E}}\left[ {{X_m}} \right] + \frac{{{\mathbb{E}}\left[ {{X_m}} \right]}}{{{{\left( {{\mathbb{E}}\left[ {{Y_m}} \right]} \right)}^2}}}{\mathbb{V}}\left[ {{Y_m}} \right] - 2\frac{{{\rm{Cov}}({X_m},{Y_m})}}{{{\mathbb{E}}\left[ {{Y_m}} \right]}}}}}_{{\xi _m}({\bf{t}})}} \right)
\end{array} \right).
\end{split}
\end{equation}
\end{figure*}

According to the above analysis, we now can obtain the closed-form ${\rm{Cov}}\left( {{X_m},{Y_m}} \right)$ as shown in (28).

Therefore, based on (9) and (13), the tight approximation about the mean and variance of $Z_m$ can be derived as
\begin{equation}
\setcounter{equation}{29}
\begin{split}{}
{\mathbb{E}}\left[ {{Z_m}} \right] \approx& {P_m}\left( {\frac{{{\mathbb{E}}\left[ {{X_m}} \right]}}{{{\mathbb{E}}\left[ {{Y_m}} \right]}} + \frac{{{\mathbb{E}}\left[ {{X_m}} \right]}}{{{{\left( {{\mathbb{E}}\left[ {{Y_m}} \right]} \right)}^3}}}{\mathbb{V}}\left[ {{Y_m}} \right]} \right.\\
&\left. { - \frac{{{\rm{Cov}}({X_m},{Y_m})}}{{{{\left( {{\mathbb{E}}\left[ {{Y_m}} \right]} \right)}^2}}}} \right),\\
{\mathbb{V}}\left[ {{Z_m}} \right] \approx &P_m^2\left( {\frac{{{\mathbb{V}}\left[ {{X_m}} \right]}}{{{{\left( {{\mathbb{E}}\left[ {{Y_m}} \right]} \right)}^2}}} + \frac{{{{\left( {{\mathbb{E}}\left[ {{X_m}} \right]} \right)}^2}}}{{{{\left( {{\mathbb{E}}\left[ {{Y_m}} \right]} \right)}^4}}}{\mathbb{V}}\left[ {{Y_m}} \right]} \right.\\
&\left. { - \frac{{2{\mathbb{E}}\left[ {{X_m}} \right]{\rm{Cov}}({X_m},{Y_m})}}{{{{\left( {{\mathbb{E}}\left[ {{Y_m}} \right]} \right)}^3}}}} \right).
\end{split}
\end{equation}

Based on (29), the PDF of $Z_m$ can be tightly approximated as a Gamma distribution by exploiting the Laguerre series approximation \cite{GUOJIE_TCOM}, i.e.,
\begin{equation}
\begin{split}{}
{f_{{Z_m}}}(v) \approx \frac{1}{{\Gamma ({\varphi _m})\vartheta _m^{{\varphi _m}}}}{v^{{\varphi _m} - 1}}{e^{ - \frac{v}{{{\vartheta _m}}}}},
\end{split}
\end{equation}
where ${\varphi _m} \buildrel \Delta \over = \frac{{{{\left( {{\mathbb{E}}\left[ {{Z_m}} \right]} \right)}^2}}}{{{\mathbb{V}}\left[ {{Z_m}} \right]}}$ is the shape parameter and ${\vartheta _m} \buildrel \Delta \over = \frac{{{\mathbb{V}}\left[ {{Z_m}} \right]}}{{{\mathbb{E}}\left[ {{Z_m}} \right]}}$ is the scale parameter.

According to the above analysis, the CDF of $Z_m$ or ${\gamma _m}({\bf{t}},{{\bf{W}}^{{\rm{ZF}}}}({\bf{t}}))$ can be approximated as
\begin{equation}
\begin{split}{}
{F_{{\gamma _m}({\bf{t}},{{\bf{W}}^{{\rm{ZF}}}}({\bf{t}}))}}\left( v \right) \approx \int_0^v {{f_{{Z_m}}}(t)} dt = \gamma \left( {{\varphi _m},\frac{v}{{{\vartheta _m}}}} \right),
\end{split}
\end{equation}
where $\gamma \left( {a,x} \right) = \frac{1}{{\Gamma (a)}}\int_0^x {{t^{a - 1}}{e^{ - t}}dt} $ is the lower incomplete gamma function.

\begin{figure}
\centering
\includegraphics[width=8cm]{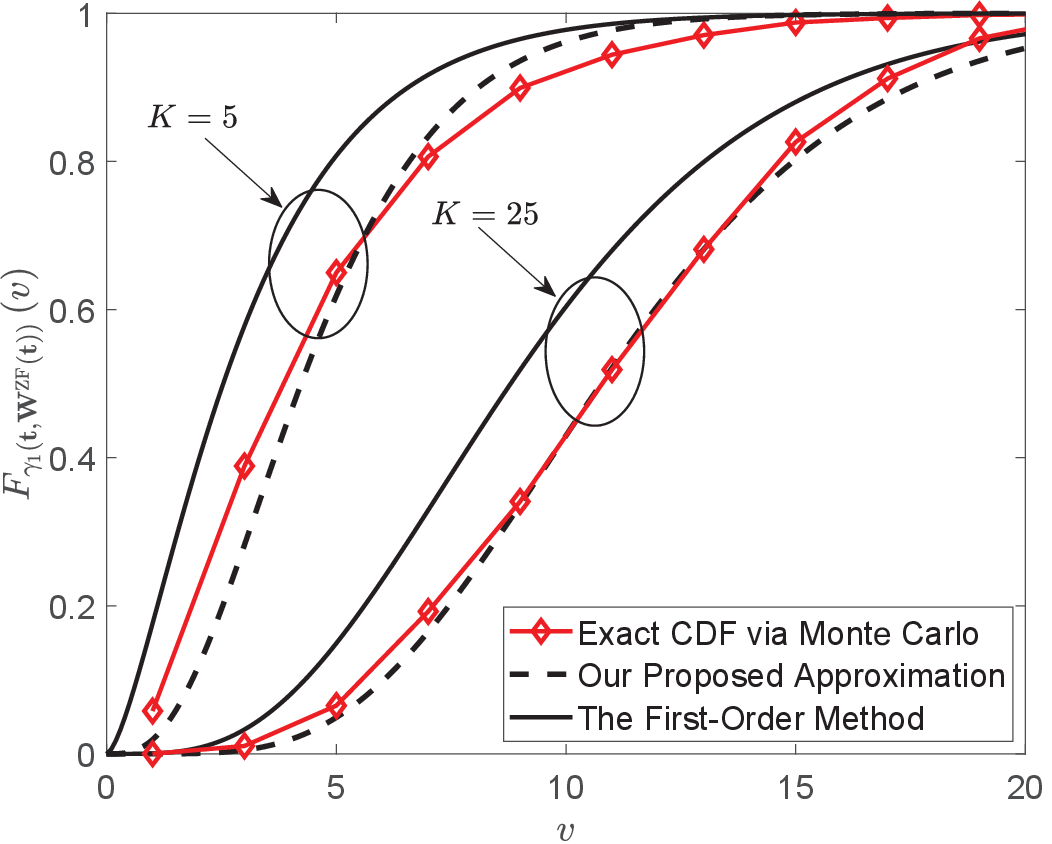}
\captionsetup{font=small}
\caption{Comparison of the CDF of ${\gamma _1}({\bf{t}},{{\bf{W}}^{{\rm{ZF}}}}({\bf{t}}))$ via Monte Carlo, the derived approximation via (31) and the first-order method via (32), where $N = 5$, $M = 4$, ${{\bf{t}}_1} = {[0,0]^T}$, ${{\bf{t}}_2} = {[0,0.5\lambda ]^T}$, ${{\bf{t}}_3} = {[0.5\lambda ,0]^T}$, ${{\bf{t}}_4} = {[0.5\lambda ,0.5\lambda ]^T}$, ${{\bf{t}}_5} = {[\lambda ,0]^T}$, ${\theta _1} = {\phi _1} = 0$, ${\theta _2} = {\phi _2} = 0.5$, ${\theta _3} = {\phi _3} = 1$, ${\theta _4} = {\phi _4} = 1.5$, $\left\{ {{P_m}} \right\}_{m = 1}^M = 10$ dBm, $\left\{ {{\beta _m}} \right\}_{m = 1}^M = {10^{ - 9}}$, ${\sigma ^2} = {10^{ - 9}}$ dBm and $\left\{ {{K_m}} \right\}_{m = 1}^M = K$.} \label{fig:Fig1}
\vspace{-10pt}
\end{figure}

\textbf{Remark 2:} If the mean of the variable $Z$ is approximated via the first-order method as in \cite{HUAMENG}, the corresponding result would become ${\left( {{\mathbb{E}}\left[ {{Z_m}} \right]} \right)_{{\rm{first}} - {\rm{order}}}} \approx {P_m}\frac{{{\mathbb{E}}\left[ {{X_m}} \right]}}{{{\mathbb{E}}\left[ {{Y_m}} \right]}}$. Based on ${\left( {{\mathbb{E}}\left[ {{Z_m}} \right]} \right)_{{\rm{first}} - {\rm{order}}}}$ and ${\mathbb{V}}\left[ {{Z_m}} \right]$, we can obtain the approximated CDF of ${\gamma _m}({\bf{t}},{{\bf{W}}^{{\rm{ZF}}}}({\bf{t}}))$ via the first-order method as
\begin{equation}
\begin{split}{}
F_{{\gamma _m}({\bf{t}},{{\bf{W}}^{{\rm{ZF}}}}({\bf{t}}))}^{{\rm{first}} - {\rm{order}}}\left( v \right) \approx &\int_0^v {{f_{{Z_m}}}(t)} dt\\
 =& \gamma \left( {\varphi _m^{{\rm{first}} - {\rm{order}}},\frac{v}{{\vartheta _m^{{\rm{first}} - {\rm{order}}}}}} \right),
\end{split}
\end{equation}
with
\begin{equation} \nonumber
\begin{split}{}
\varphi _m^{{\rm{first}} - {\rm{order}}} \buildrel \Delta \over = &\frac{{{{\left( {{{\left( {{\mathbb{E}}\left[ {{Z_m}} \right]} \right)}_{{\rm{first}} - {\rm{order}}}}} \right)}^2}}}{{{\mathbb{V}}\left[ {{Z_m}} \right]}},\\
\vartheta _m^{{\rm{first}} - {\rm{order}}} = &\frac{{{\mathbb{V}}\left[ {{Z_m}} \right]}}{{{{\left( {{\mathbb{E}}\left[ {{Z_m}} \right]} \right)}_{{\rm{first}} - {\rm{order}}}}}}.
\end{split}
\end{equation}

  \begin{figure*}[b!]
  \hrulefill
\setcounter{mytempeqncnt}{\value{equation}}
\setcounter{equation}{38}
\begin{equation}
\begin{split}{}
{R_m} \approx R_m^{{\rm{appro}}} =& {\log _2}\left( {\underbrace {{\varpi _{m,1}} + {\varpi _{m,2}}{f_{m,1}}({\bf{t}}) + {f_{m,2}}({\bf{t}})\left( {{\varpi _{m,3}} + {\varpi _{m,4}}{f_{m,1}}({\bf{t}})} \right) + \sum\nolimits_{j = 1,j \ne m}^M {{\varpi _{mj,5}}{f_{mj,3}}({\bf{t}})} }_{{f_{m,4}}({\bf{t}})}} \right.\\
 &+ \left. {\frac{{\overbrace {{\varpi _{m,6}} + {f_{m,2}}({\bf{t}})\left( {{\varpi _{m,7}} + {\varpi _{m,8}}{f_{m,1}}({\bf{t}})} \right) + \sum\nolimits_{j = 1,j \ne m}^M {{\varpi _{mj,9}}{f_{mj,3}}({\bf{t}})} }^{{f_{m,5}}({\bf{t}})}}}{{\underbrace {{\varpi _{m,10}} + {\varpi _{m,11}}{f_{m,2}}({\bf{t}}) + \frac{{\sum\nolimits_{j = 1,j \ne m}^M {{\varpi _{mj,12}}{f_{mj,3}}({\bf{t}})} }}{{{\varpi _{m,13}}{f_{m,1}}({\bf{t}}) + {\varpi _{m,14}}}}}_{{f_{m,6}}({\bf{t}})}}}} \right).
\end{split}
\end{equation}
\end{figure*}

\setcounter{equation}{\value{mytempeqncnt}}

We plot the CDF curves of ${\gamma _m}({\bf{t}},{{\bf{W}}^{{\rm{ZF}}}}({\bf{t}}))$ via Monte Carlo, our proposed approximation and the first-order method as shown in Fig. 2, from which it is clear that our proposed approximation is closer to Monte Carlo. Therefore, our proposed approximation is accurate enough to support the subsequent analysis.

\section{The Approximated Expression of The Outage-Aware Rate}
Based on the closed-form CDF of ${\gamma _m}({\bf{t}},{{\bf{W}}^{{\rm{ZF}}}}({\bf{t}}))$ presented in (31), we now can replace the constraint in (6d) as
\begin{equation}
\setcounter{equation}{33}
\begin{split}{}
&\Pr \left\{ {{\gamma _m}({\bf{t}},{{\bf{W}}^{{\rm{ZF}}}}({\bf{t}})) < {2^{{R_m}}} - 1} \right\} = \delta \\
 \Leftrightarrow &\gamma \left( {{\varphi _m},\frac{{{2^{{R_m}}} - 1}}{{{\vartheta _m}}}} \right) = \delta \\
 \Leftrightarrow &{R_m} = {\log _2}\left( {1 + {\vartheta _m}{\gamma ^{ - 1}}\left( {\delta ,{\varphi _m}} \right)} \right),
\end{split}
\end{equation}
where ${{\gamma ^{ - 1}}\left( {\delta ,{\varphi _m}} \right)}$, with $\delta  \in [0,1]$, is the inverse of the lower incomplete gamma function.

Although we have obtained the closed-form outage-aware rate in (33), such expression includes the complex ${{\gamma ^{ - 1}}\left( {\delta ,{\varphi _m}} \right)}$ which greatly hinders the subsequent optimization analysis. Motivated by this challenge, given $\delta $, we now exploit the conclusion in our previous work \cite{Guojie1} to effectively approximate ${{\gamma ^{ - 1}}\left( {\delta ,{\varphi _m}} \right)}$ as a linear function with respect to ${\varphi _m}$, i.e.,
\begin{equation}
\begin{split}{}
{\gamma ^{ - 1}}\left( {\delta ,{\varphi _m}} \right) \approx \kappa (\delta ){\varphi _m} + \rho (\delta ),
\end{split}
\end{equation}
where $\kappa (\delta )$ and $\rho (\delta )$ are the parameters related to $\delta $ and are determined based on the criterion of minimum mean square error. Generally, $\delta $ should be set in a small range by the system to strictly control the outage behavior. Based on this fact, given $\delta  \in [0.1:0.01:0.2]$ as an example, we list here the corresponding $\kappa (\delta )$ and $\rho (\delta )$ presented in Table I, armed with which others can better verify the tightness of our proposed linear approximation shown in (34) for the inverse of the incomplete gamma function.

Based on (33) and (34), $R_m$ can be approximated as
\begin{equation}
\begin{split}{}
{R_m} \approx &{\log _2}\left( {1 + {\vartheta _m}\left( {\kappa (\delta ){\varphi _m} + \rho (\delta )} \right)} \right)\\
 =& {\log _2}\left( {1 + \kappa (\delta ){\vartheta _m}{\varphi _m} + \rho (\delta ){\vartheta _m}} \right)\\
 =& {\log _2}\left( {1 + \kappa (\delta ){\mathbb{E}}\left[ {{Z_m}} \right] + \rho (\delta )\frac{{{\mathbb{V}}\left[ {{Z_m}} \right]}}{{{\mathbb{E}}\left[ {{Z_m}} \right]}}} \right),
\end{split}
\end{equation}
substituting the expressions of ${{\mathbb{E}}\left[ {{Z_m}} \right]}$ and ${{\mathbb{V}}\left[ {{Z_m}} \right]}$ into (35), $R_m$ can be expanded as in (36). In addition, since generally ${K_m}{\left| {{{\overline {\bf{h}} }_m}({\bf{t}}){\bf{w}}_m^{{\rm{ZF}}}({\bf{t}})} \right|^2} \gg 1$ especially for the large $K_m$ and $N$, we can obtain the following tight approximation
\begin{equation}
\setcounter{equation}{37}
\begin{split}{}
\frac{{{\mathbb{V}}\left[ {{X_m}} \right]}}{{{\mathbb{E}}\left[ {{X_m}} \right]}} =& \frac{{\frac{{{\beta _m}}}{{{K_m} + 1}}\left( {1 + 2{K_m}{{\left| {{{\overline {\bf{h}} }_m}({\bf{t}}){\bf{w}}_m^{{\rm{ZF}}}({\bf{t}})} \right|}^2}} \right)}}{{1 + {K_m}{{\left| {{{\overline {\bf{h}} }_m}({\bf{t}}){\bf{w}}_m^{{\rm{ZF}}}({\bf{t}})} \right|}^2}}}\\
 \approx & \frac{{2{\beta _m}}}{{{K_m} + 1}},
\end{split}
\end{equation}
based on which ${\xi _m}({\bf{t}})$ in (36) can be approximated as
\begin{equation}
\begin{split}{}
{\xi _m}({\bf{t}}) \approx \frac{{\frac{{2{\beta _m}}}{{{K_m} + 1}} + \frac{{{\mathbb{E}}\left[ {{X_m}} \right]}}{{{{\left( {{\mathbb{E}}\left[ {{Y_m}} \right]} \right)}^2}}}{\mathbb{V}}\left[ {{Y_m}} \right] - 2\frac{{{\rm{Cov}}({X_m},{Y_m})}}{{{\mathbb{E}}\left[ {{Y_m}} \right]}}}}{{1 + \frac{{{\mathbb{V}}\left[ {{Y_m}} \right]}}{{{{\left( {{\mathbb{E}}\left[ {{Y_m}} \right]} \right)}^2}}} - 2\frac{{{\rm{Cov}}({X_m},{Y_m})}}{{{\mathbb{E}}\left[ {{X_m}} \right]{\mathbb{E}}\left[ {{Y_m}} \right]}}}}.
\end{split}
\end{equation}

According to the above analysis and substituting the known expressions of ${{\mathbb{E}}\left[ {{X_m}} \right]}$, ${{\mathbb{E}}\left[ {{Y_m}} \right]}$, ${{\mathbb{V}}\left[ {{X_m}} \right]}$, ${{\mathbb{V}}\left[ {{Y_m}} \right]}$, ${{\rm{Cov}}({X_m},{Y_m})}$ and the approximated ${\xi _m}({\bf{t}})$ in (38) into (36), we can simplify $R_m$ as in (39), where
\begin{equation} \nonumber
\begin{split}{}
{\varpi _{m,1}} \buildrel \Delta \over =& 1 + \frac{{\kappa (\delta ){P_m}{\beta _m}}}{{{\mathbb{E}}\left[ {{Y_m}} \right]\left( {{K_m} + 1} \right)}},{\varpi _{m,2}} \buildrel \Delta \over = \frac{{\kappa (\delta ){P_m}{\beta _m}{K_m}}}{{{\mathbb{E}}\left[ {{Y_m}} \right]\left( {{K_m} + 1} \right)}},\\
{\varpi _{m,3}} \buildrel \Delta \over =& \frac{{\kappa (\delta ){P_m}\beta _m^3}}{{{{\left( {{\mathbb{E}}\left[ {{Y_m}} \right]} \right)}^3}{{\left( {{K_m} + 1} \right)}^3}}},{\varpi _{m,4}} \buildrel \Delta \over = \frac{{\kappa (\delta ){P_m}{K_m}\beta _m^3}}{{{{\left( {{\mathbb{E}}\left[ {{Y_m}} \right]} \right)}^3}{{\left( {{K_m} + 1} \right)}^3}}},\\
{\varpi _{mj,5}} \buildrel \Delta \over =&  - \frac{{\kappa (\delta ){P_j}{P_m}\beta _m^2}}{{{{\left( {{\mathbb{E}}\left[ {{Y_m}} \right]} \right)}^2}{{\left( {{K_m} + 1} \right)}^2}}},{\varpi _{m,6}} \buildrel \Delta \over = \frac{{2\rho (\delta ){P_m}{\beta _m}}}{{{K_m} + 1}},\\
{\varpi _{m,7}} \buildrel \Delta \over =& \frac{{\rho (\delta ){P_m}\beta _m^3}}{{{{\left( {{\mathbb{E}}\left[ {{Y_m}} \right]} \right)}^2}{{\left( {{K_m} + 1} \right)}^3}}},{\varpi _{m,8}} \buildrel \Delta \over = \frac{{\rho (\delta ){P_m}{K_m}\beta _m^3}}{{{{\left( {{\mathbb{E}}\left[ {{Y_m}} \right]} \right)}^2}{{\left( {{K_m} + 1} \right)}^3}}},
\end{split}
\end{equation}
\begin{equation} \nonumber
\begin{split}{}
{\varpi _{mj,9}} \buildrel \Delta \over =& \frac{{ - 2\rho (\delta ){P_m}{P_j}\beta _m^2}}{{{\mathbb{E}}\left[ {{Y_m}} \right]{{\left( {{K_m} + 1} \right)}^2}}},{\varpi _{m,10}} \buildrel \Delta \over = {\mathbb{E}}\left[ {{Y_m}} \right],\\
{\varpi _{m,11}} \buildrel \Delta \over =& \frac{{\beta _m^2}}{{{\mathbb{E}}\left[ {{Y_m}} \right]{{\left( {{K_m} + 1} \right)}^2}}},{\varpi _{mj,12}} \buildrel \Delta \over = \frac{{ - 2{P_j}\beta _m^2}}{{{{\left( {{K_m} + 1} \right)}^2}}},\\
{\varpi _{m,13}} \buildrel \Delta \over =& \frac{{{K_m}{\beta _m}}}{{{K_m} + 1}},{\varpi _{m,14}} \buildrel \Delta \over = \frac{{{\beta _m}}}{{{K_m} + 1}}
\end{split}
\end{equation}
and
\begin{equation} \nonumber
\begin{split}{}
{f_{m,1}}({\bf{t}}) \buildrel \Delta \over =& {\left| {{{\overline {\bf{h}} }_m}({\bf{t}}){\bf{w}}_m^{{\rm{ZF}}}({\bf{t}})} \right|^2},\\
{f_{m,2}}({\bf{t}}) \buildrel \Delta \over =& {\rm{tr}}\left( {{{\left( {{{\bf{\Psi }}_m}({\bf{t}})} \right)}^2}} \right),\\
{f_{mj,3}}({\bf{t}}) \buildrel \Delta \over =& {\left| {{{\left( {{\bf{w}}_j^{{\rm{ZF}}}({\bf{t}})} \right)}^H}{\bf{w}}_m^{{\rm{ZF}}}({\bf{t}})} \right|^2}.
\end{split}
\end{equation}

\begin{figure}
\centering
\includegraphics[width=8cm]{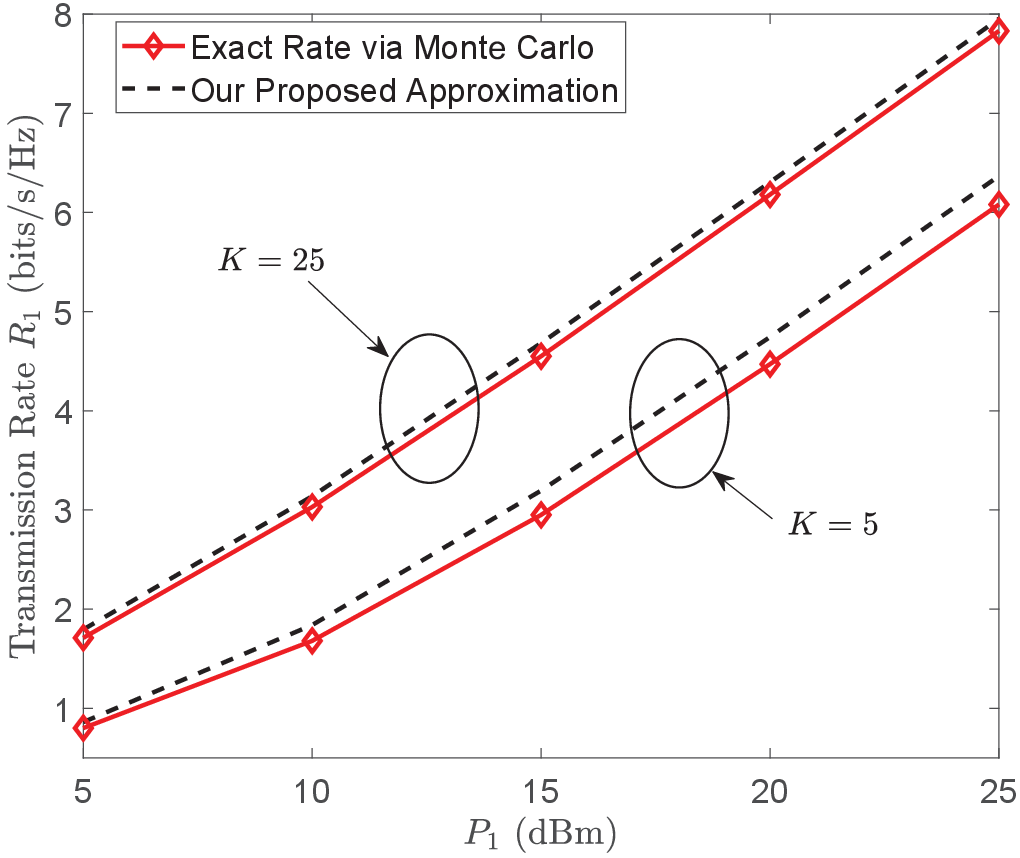}
\captionsetup{font=small}
\caption{Comparison of the outage-aware transmission rate of the first user via Monte Carlo and our proposed approximation via (39), where $\left\{ {{P_m}} \right\}_{m = 2}^M = 10$ dBm, the target outage probability is set as $\delta  = 0.2$, and other parameters are the same as those in Fig. 2.} \label{fig:Fig2}
\vspace{-10pt}
\end{figure}

\textbf{Remark 3:} Note that the linear function is designed to approximate the inverse of the lower incomplete gamma function as shown in (34), and concurrently, $\frac{{2{\beta _m}}}{{{K_m} + 1}}$ is provided to approximate $\frac{{{\mathbb{V}}\left[ {{X_m}} \right]}}{{{\mathbb{E}}\left[ {{X_m}} \right]}}$. Although both of these approximations are tight, they may still introduce some error between the approximated rate and the exact one. Therefore, we have to compare the outage-aware transmission rate via Monte Carlo and our proposed approximation via (39) as presented in Fig. 3, from which it can be verified that our proposed approximation is almost close to the exact one. Based on this fact, in the next section, it is reasonable to perform optimization analysis based on the approximated rate in (39).

\section{Optimizing Antenna Positions Via PGA}
According to the above analysis, we now can simplify problem (P1) as
 \begin{align}
&({\rm{P2}}):{\rm{  }}\mathop {\max }\limits_{{\bf{t}}} \ \sum\nolimits_{m = 1}^M {R_m^{{\rm{appro}}}}  \tag{${\rm{40a}}$}\\
{\rm{              }}&{\rm{s.t.}} \ \ \ {{\bf{t}}_n} \in {{\cal A}_n},\forall n \in {\cal N}.\tag{${\rm{40b}}$}
 \end{align}

Generally, (P2) is highly non-convex since its objective includes complex structures. However, we note that each variable ${{\bf{t}}_n}$ is located within a constant range ${{\cal A}_n}$ in the constraint (40b). This motivates us to employ the PGA method for solving (P2), since the PGA is well-suited for problems with simple constraints and is not sensitive to concavity or convexity of the objective.

With the PGA method, the update rule for ${{\bf{t}}}$ in the $k + 1$-th iteration is given by
\begin{equation}
\setcounter{equation}{41}
\begin{split}{}
{{\bf{t}}^{(k + 1)}} =& {{\bf{t}}^{(k)}} + \varsigma {\nabla _{{{\bf{t}}^{(k)}}}}\sum\nolimits_{m = 1}^M {R_m^{{\rm{appro}}}} ,\\
{{\bf{t}}^{(k + 1)}} =& {\cal B}\left\{ {{{\bf{t}}^{(k + 1)}},\left\{ {{{\cal A}_n}} \right\}_{n = 1}^N} \right\},
\end{split}
\end{equation}
where ${\nabla _{{{\bf{t}}^{(k)}}}}\sum\nolimits_{m = 1}^M {R_m^{{\rm{appro}}}}  \in {{\mathbb{R}}^{2 \times N}}$ denotes the gradient of $\sum\nolimits_{m = 1}^M {R_m^{{\rm{appro}}}} $ at ${{\bf{t}}^{(k)}}$, and $\varsigma $ is the step size for the gradient ascent, which can be set properly based on the backtracking line search []. In addition, ${\cal B}\left\{  \cdot  \right\}$ is the projection function, which ensures that the solutions for MAs' positions in each iteration always satisfy the constraint in (40b), i.e.,
\begin{equation} \nonumber
\begin{split}{}
&{{\bf{t}}^{(k + 1)}} = {\cal B}\left\{ {{{\bf{t}}^{(k + 1)}},\left\{ {{{\cal A}_n}} \right\}_{n = 1}^N} \right\}\\
 \Leftrightarrow & \left\{ {\begin{array}{*{20}{c}}
{x_n^{(k + 1)} = \min \left( {\max \left( {x_n^{\min },x_n^{(k + 1)}} \right),x_n^{\max }} \right)}\\
{y_n^{(k + 1)} = \min \left( {\max \left( {y_n^{\min },x_n^{(k + 1)}} \right),y_n^{\max }} \right)}
\end{array}} \right.,\forall n \in {\cal N}.
\end{split}
\end{equation}

 Observing the above process, to successfully run the PGA algorithm, the key is to derive the expression of ${\nabla _{{{\bf{t}}^{(k)}}}}\sum\nolimits_{m = 1}^M {R_m^{{\rm{appro}}}} $. To proceed, note that
\begin{equation}
\begin{split}{}
&{\nabla _{{{\bf{t}}^{(k)}}}}\sum\nolimits_{m = 1}^M {R_m^{{\rm{appro}}}}  \\
=& {\left[ {\frac{{\partial \sum\nolimits_{m = 1}^M {R_m^{{\rm{appro}}}} }}{{\partial {{\bf{t}}_1}}},...,\frac{{\partial \sum\nolimits_{m = 1}^M {R_m^{{\rm{appro}}}} }}{{\partial {{\bf{t}}_N}}}} \right]_{{\bf{t}} = {{\bf{t}}^{(k)}}}},
\end{split}
\end{equation}
with
\begin{equation} \nonumber
\begin{split}{}
\frac{{\partial \sum\nolimits_{m = 1}^M {R_m^{{\rm{appro}}}} }}{{\partial {{\bf{t}}_n}}} = {\left[ {\frac{{\partial \sum\nolimits_{m = 1}^M {R_m^{{\rm{appro}}}} }}{{\partial {x_n}}},\frac{{\partial \sum\nolimits_{m = 1}^M {R_m^{{\rm{appro}}}} }}{{\partial {y_n}}}} \right]^T}.
\end{split}
\end{equation}

In the next, we mainly aim to derive the closed-form ${\frac{{\partial \sum\nolimits_{m = 1}^M {R_m^{{\rm{appro}}}} }}{{\partial {x_n}}}}$, $\forall n \in {\cal N}$, and the processes of deriving ${\frac{{\partial \sum\nolimits_{m = 1}^M {R_m^{{\rm{appro}}}} }}{{\partial {y_n}}}}$ are similar to the above, which thus are omitted here for brevity.

Specifically, based on (39), ${\frac{{\partial \sum\nolimits_{m = 1}^M {R_m^{{\rm{appro}}}} }}{{\partial {x_n}}}}$ is computed as
\begin{equation}
\begin{split}{}
&\frac{{\partial \sum\nolimits_{m = 1}^M {R_m^{{\rm{appro}}}} }}{{\partial {x_n}}} = \frac{{\partial \sum\nolimits_{m = 1}^M {{{\log }_2}\left( {{f_{m,4}}({\bf{t}}) + \frac{{{f_{m,5}}({\bf{t}})}}{{{f_{m,6}}({\bf{t}})}}} \right)} }}{{\partial {x_n}}}\\
 =& \frac{1}{{\ln 2}}\sum\nolimits_{m = 1}^M {\frac{1}{{{f_{m,4}}({\bf{t}}) + \frac{{{f_{m,5}}({\bf{t}})}}{{{f_{m,6}}({\bf{t}})}}}}} \\
& \times \left( {\frac{{\partial {f_{m,4}}({\bf{t}})}}{{\partial {x_n}}} + \frac{{\frac{{\partial {f_{m,5}}({\bf{t}})}}{{\partial {x_n}}}{f_{m,6}}({\bf{t}}) - {f_{m,5}}({\bf{t}})\frac{{\partial {f_{m,6}}({\bf{t}})}}{{\partial {x_n}}}}}{{{{\left( {\frac{{{f_{m,5}}({\bf{t}})}}{{{f_{m,6}}({\bf{t}})}}} \right)}^2}}}} \right),
\end{split}
\end{equation}
where
\begin{equation} \nonumber
\begin{split}{}
&\frac{{\partial {f_{m,4}}({\bf{t}})}}{{\partial {x_n}}} = {\varpi _{m,2}}\frac{{\partial {f_{m,1}}({\bf{t}})}}{{\partial {x_n}}} + \frac{{\partial {f_{m,2}}({\bf{t}})}}{{\partial {x_n}}}\\
& \times \left( {{\varpi _{m,3}} + {\varpi _{m,4}}{f_{m,1}}({\bf{t}})} \right) + {\varpi _{m,4}}{f_{m,2}}({\bf{t}})\\
 &\times \frac{{\partial {f_{m,1}}({\bf{t}})}}{{\partial {x_n}}} + \sum\nolimits_{j = 1,j \ne m}^M {{\varpi _{mj,5}}\frac{{\partial {f_{mj,3}}({\bf{t}})}}{{\partial {x_n}}}} ,\\
&\frac{{\partial {f_{m,5}}({\bf{t}})}}{{\partial {x_n}}} = \frac{{\partial {f_{m,2}}({\bf{t}})}}{{\partial {x_n}}}\left( {{\varpi _{m,7}} + {\varpi _{m,8}}{f_{m,1}}({\bf{t}})} \right)\\
& + {\varpi _{m,8}}{f_{m,2}}({\bf{t}})\frac{{\partial {f_{m,1}}({\bf{t}})}}{{\partial {x_n}}} + \sum\nolimits_{j = 1,j \ne m}^M {{\varpi _{mj,9}}\frac{{\partial {f_{mj,3}}({\bf{t}})}}{{\partial {x_n}}}}, \\
&\frac{{\partial {f_{m,6}}({\bf{t}})}}{{\partial {x_n}}} = {\varpi _{m,11}}\frac{{\partial {f_{m,2}}({\bf{t}})}}{{\partial {x_n}}}\\
 &+ \frac{{\sum\nolimits_{j = 1,j \ne m}^M {{\varpi _{mj,12}}\frac{{\partial {f_{mj,3}}({\bf{t}})}}{{\partial {x_n}}}\left( {{\varpi _{m,13}}{f_{m,1}}({\bf{t}}) + {\varpi _{m,14}}} \right)} }}{{{{\left( {{\varpi _{m,13}}{f_{m,1}}({\bf{t}}) + {\varpi _{m,14}}} \right)}^2}}}\\
 &- \frac{{\sum\nolimits_{j = 1,j \ne m}^M {{\varpi _{mj,12}}{f_{mj,3}}({\bf{t}}){\varpi _{m,13}}\frac{{\partial {f_{m,1}}({\bf{t}})}}{{\partial {x_n}}}} }}{{{{\left( {{\varpi _{m,13}}{f_{m,1}}({\bf{t}}) + {\varpi _{m,14}}} \right)}^2}}}.
\end{split}
\end{equation}

Note that $\frac{{\partial {f_{m,4}}({\bf{t}})}}{{\partial {x_n}}}$, $\frac{{\partial {f_{m,5}}({\bf{t}})}}{{\partial {x_n}}}$ and $\frac{{\partial {f_{m,6}}({\bf{t}})}}{{\partial {x_n}}}$ are all related to $\frac{{\partial {f_{m,1}}({\bf{t}})}}{{\partial {x_n}}}$, $\frac{{\partial {f_{m,2}}({\bf{t}})}}{{\partial {x_n}}}$ and $\left\{ {\frac{{\partial {f_{mj,3}}({\bf{t}})}}{{\partial {x_n}}}} \right\}_{j = 1,j \ne m}^M$, the closed-form expressions of which are provided in the follows.

1) First, note that
\begin{equation}
\begin{split}{}
{f_{m,1}}({\bf{t}}) \buildrel \Delta \over =& {\left| {{{\overline {\bf{h}} }_m}({\bf{t}}){\bf{w}}_m^{{\rm{ZF}}}({\bf{t}})} \right|^2} = \frac{1}{{{{\left[ {\overline {{\bf{HH}}} ({\bf{t}})} \right]}_{mm}}}},
\end{split}
\end{equation}
where $\overline {{\bf{HH}}} ({\bf{t}}) \buildrel \Delta \over = {\left( {{{\overline {\bf{H}} }^H}({\bf{t}})\overline {\bf{H}} ({\bf{t}})} \right)^{ - 1}}$. Based on (44) we have
\begin{equation}
\begin{split}{}
\frac{{\partial {f_{m,1}}({\bf{t}})}}{{\partial {x_n}}} =&  - \frac{{\partial {{\left[ {\overline {{\bf{HH}}} ({\bf{t}})} \right]}_{mm}}/\partial {x_n}}}{{{{\left( {{{\left[ {\overline {{\bf{HH}}} ({\bf{t}})} \right]}_{mm}}} \right)}^2}}}\\
 =&  - \frac{{{{\left[ {\partial \left[ {\overline {{\bf{HH}}} ({\bf{t}})} \right]/\partial {x_n}} \right]}_{mm}}}}{{{{\left( {{{\left[ {\overline {{\bf{HH}}} ({\bf{t}})} \right]}_{mm}}} \right)}^2}}},
\end{split}
\end{equation}
where
\begin{equation} \nonumber
\begin{split}{}
\partial \left[ {\overline {{\bf{HH}}} ({\bf{t}})} \right]/\partial {x_n} =&  - \overline {{\bf{HH}}} ({\bf{t}})\\
 &\times \partial \left[ {{{\overline {\bf{H}} }^H}({\bf{t}})\overline {\bf{H}} ({\bf{t}})} \right]/\partial {x_n}\overline {{\bf{HH}}} ({\bf{t}}),
\end{split}
\end{equation}
with
\begin{equation} \nonumber
\begin{split}{}
\frac{{\partial \left[ {{{\overline {\bf{H}} }^H}({\bf{t}})\overline {\bf{H}} ({\bf{t}})} \right]}}{{\partial {x_n}}} = \frac{{\partial {{\overline {\bf{H}} }^H}({\bf{t}})}}{{\partial {x_n}}}\overline {\bf{H}} ({\bf{t}}) + {\overline {\bf{H}} ^H}({\bf{t}})\frac{{\partial \overline {\bf{H}} ({\bf{t}})}}{{\partial {x_n}}},
\end{split}
\end{equation}
and $\frac{{\partial \overline {\bf{H}} ({\bf{t}})}}{{\partial {x_n}}} = {\left( {\frac{{\partial {{\overline {\bf{H}} }^H}({\bf{t}})}}{{\partial {x_n}}}} \right)^H}$, with
\begin{equation} \nonumber
\begin{split}{}
\frac{{\partial {{\overline {\bf{H}} }^H}({\bf{t}})}}{{\partial {x_n}}} = \left[ {{{\bf{0}}^{M \times 1}},...,{\bf{b}}({x_n}),...,{{\bf{0}}^{M \times 1}}} \right] \in {{\mathbb{C}}^{M \times N}},
\end{split}
\end{equation}
where ${\bf{b}}({x_n}) \in {{\mathbb{C}}^{M \times 1}}$ is the $n$-th column vector of $\frac{{\partial {{\overline {\bf{H}} }^H}({\bf{t}})}}{{\partial {x_n}}}$, with
\begin{equation} \nonumber
\begin{split}{}
&{\bf{b}}({x_n}) = \frac{{2\pi }}{\lambda }\\
 \times& \left[ {\cos {\theta _m}\sin {\phi _m}{e^{j\left( {\frac{{2\pi }}{\lambda }\left( {{x_n}\cos {\theta _m}\sin {\phi _m} + {y_n}\sin {\theta _m}} \right) + \frac{\pi }{2}} \right)}}} \right]_{m = 1:1:M}^T.
\end{split}
\end{equation}

To this end, combining the above results yields the closed-form expression for $\frac{{\partial {f_{m,1}}({\bf{t}})}}{{\partial {x_n}}}$.

2) Second, note that
\begin{equation}
\begin{split}{}
&{f_{m,2}}({\bf{t}}) \buildrel \Delta \over = {\rm{tr}}\left( {{{\left( {\sum\nolimits_{j = 1,j \ne m}^M {{P_j}{\bf{w}}_j^{{\rm{ZF}}}({\bf{t}}){{\left( {{\bf{w}}_j^{{\rm{ZF}}}({\bf{t}})} \right)}^H}} } \right)}^2}} \right)\\
 =& \sum\nolimits_{i = 1,i \ne m}^M {\sum\nolimits_{j = 1,j \ne m}^M {{P_i}{P_j}} } \frac{{{{\left| {{{\left[ {\overline {{\bf{HH}}} ({\bf{t}})} \right]}_{ij}}} \right|}^2}}}{{{{\left[ {\overline {{\bf{HH}}} ({\bf{t}})} \right]}_{ii}}{{\left[ {\overline {{\bf{HH}}} ({\bf{t}})} \right]}_{jj}}}},
\end{split}
\end{equation}
based on which $\frac{{\partial {f_{m,2}}({\bf{t}})}}{{\partial {x_n}}}$ can be derived as
\begin{equation}
\begin{split}{}
&\frac{{\partial {f_{m,2}}({\bf{t}})}}{{\partial {x_n}}} = \sum\nolimits_{i = 1,i \ne m}^M {\sum\nolimits_{j = 1,j \ne m}^M {{P_i}{P_j}} } \\
 \times& \left( {\frac{{\partial {{\left| {{{\left[ {\overline {{\bf{HH}}} ({\bf{t}})} \right]}_{ij}}} \right|}^2}/\partial {x_n}}}{{{{\left[ {\overline {{\bf{HH}}} ({\bf{t}})} \right]}_{ii}}{{\left[ {\overline {{\bf{HH}}} ({\bf{t}})} \right]}_{jj}}}}} \right.\\
 &- \frac{{{{\left| {{{\left[ {\overline {{\bf{HH}}} ({\bf{t}})} \right]}_{ij}}} \right|}^2}\partial {{\left[ {\overline {{\bf{HH}}} ({\bf{t}})} \right]}_{ii}}/\partial {x_n}}}{{{{\left( {{{\left[ {\overline {{\bf{HH}}} ({\bf{t}})} \right]}_{ii}}} \right)}^2}{{\left[ {\overline {{\bf{HH}}} ({\bf{t}})} \right]}_{jj}}}}\\
&\left. { - \frac{{{{\left| {{{\left[ {\overline {{\bf{HH}}} ({\bf{t}})} \right]}_{ij}}} \right|}^2}\partial {{\left[ {\overline {{\bf{HH}}} ({\bf{t}})} \right]}_{jj}}/\partial {x_n}}}{{{{\left( {{{\left[ {\overline {{\bf{HH}}} ({\bf{t}})} \right]}_{jj}}} \right)}^2}{{\left[ {\overline {{\bf{HH}}} ({\bf{t}})} \right]}_{ii}}}}} \right).
\end{split}
\end{equation}
 In (47), since $\frac{{\partial {{\left| {{{\left[ {\overline {{\bf{HH}}} ({\bf{t}})} \right]}_{ij}}} \right|}^2}}}{{\partial {x_n}}} = \frac{{\partial {{\left[ {\overline {{\bf{HH}}} ({\bf{t}})} \right]}_{ij}}}}{{\partial {x_n}}}\left[ {\overline {{\bf{HH}}} ({\bf{t}})} \right]_{ij}^H + {\left[ {\overline {{\bf{HH}}} ({\bf{t}})} \right]_{ij}}\frac{{\partial \left[ {\overline {{\bf{HH}}} ({\bf{t}})} \right]_{ij}^H}}{{\partial {x_n}}}$, where the closed-form expression of $\frac{{\partial {{\left[ {\overline {{\bf{HH}}} ({\bf{t}})} \right]}_{ij}}}}{{\partial {x_n}}} = {\left[ {\frac{{\partial {{\left( {{{\overline {\bf{H}} }^H}({\bf{t}})\overline {\bf{H}} ({\bf{t}})} \right)}^{ - 1}}}}{{\partial {x_n}}}} \right]_{ij}}$ or $\frac{{\partial \left[ {\overline {{\bf{HH}}} ({\bf{t}})} \right]_{ij}^H}}{{\partial {x_n}}}$, $\forall i,j \in {\cal M}$ and $i,j \ne m$, has been derived earlier, based on which we can directly obtain the closed-form $\frac{{\partial {f_{m,2}}({\bf{t}})}}{{\partial {x_n}}}$.

3) Third, note that
\begin{equation}
\begin{split}{}
{f_{mj,3}}({\bf{t}}) \buildrel \Delta \over = &{\left| {{{\left( {{\bf{w}}_j^{{\rm{ZF}}}({\bf{t}})} \right)}^H}{\bf{w}}_m^{{\rm{ZF}}}({\bf{t}})} \right|^2}\\
 =& \frac{{{{\left| {{{\left[ {\overline {{\bf{HH}}} ({\bf{t}})} \right]}_{jm}}} \right|}^2}}}{{{{\left[ {\overline {{\bf{HH}}} ({\bf{t}})} \right]}_{jj}}{{\left[ {\overline {{\bf{HH}}} ({\bf{t}})} \right]}_{mm}}}}.
\end{split}
\end{equation}
Since the structure of ${f_{mj,3}}({\bf{t}})$ is similar to that of ${{f_{m,2}}({\bf{t}})}$, the closed-form $\frac{{\partial {f_{mj,3}}({\bf{t}})}}{{\partial {x_n}}}$ can be derived directly based on the above results and thus the details are omitted here.

Finally, substituting $\frac{{\partial {f_{m,1}}({\bf{t}})}}{{\partial {x_n}}}$, $\frac{{\partial {f_{m,2}}({\bf{t}})}}{{\partial {x_n}}}$ and $\frac{{\partial {f_{mj,3}}({\bf{t}})}}{{\partial {x_n}}}$ into (43), the closed-form $\frac{{\partial \sum\nolimits_{m = 1}^M {R_m^{{\rm{appro}}}} }}{{\partial {x_n}}}$ can be successfully obtained.

\begin{algorithm}
\caption{The PGA for Solving (P2)}
  \begin{algorithmic}[1]

\State \textbf{{Input:}} Generate $I$ random antenna positions $\left\{ {{{\bf{t}}^{(1,i)}}} \right\}_{i = 1}^I$, where ${\bf{t}}_n^{(1,i)} \in {{\cal A}_n}$, $\forall i,n$ and set $k = 1$.

\State {\textbf{For}} $i = 1:1:I$

\State \quad Let ${{\bf{t}}^{(1)}} = {\bf{t}}_n^{(1,i)}$,

\State \textbf{Repeat:}

\State \quad ${{\bf{t}}^{(k + 1)}} = {{\bf{t}}^{(k)}} + \varsigma {\nabla _{{{\bf{t}}^{(k)}}}}\sum\nolimits_{m = 1}^M {R_m^{{\rm{appro}}}}$ ,
\State \quad ${{\bf{t}}^{(k + 1)}} = {\cal B}\left\{ {{{\bf{t}}^{(k + 1)}},\left\{ {{{\cal A}_n}} \right\}_{n = 1}^N} \right\}$,
\State \quad $k \leftarrow k + 1$.
\State \quad \textbf{Until:} $\sum\nolimits_{m = 1}^M {R_m^{{\rm{appro}}}} $ converges to a stationary value, denoted by ${\rm{O}}{{\rm{B}}_i}$.
\State {\textbf{End For}}.
\State \textbf{Output:} Select the maximum value from $\left\{ {{\rm{O}}{{\rm{B}}_i}} \right\}_{i = 1}^I$ and then the corresponding antenna position deployments.
  \end{algorithmic}
\end{algorithm}

\textbf{Remark 4:} Note that (P2) is a multivariate optimization problem, and the step-by-step ascent mechanism in the PGA method makes it highly effective at finding a local maximum but provides no guarantee that this maximum will be the global one. Therefore, a common and practical strategy to mitigate the local optimum is to run the PGA method multiple times from a set of distinct, randomly chosen initial points. Upon completion, the solution associated with the run that achieved the maximum objective value is selected as the final outcome. The process that describes the PGA method for solving (P2) is provided in Algorithm 1.

\textbf{Complexity Analysis:} We first analyze the complexity of computing $\frac{{\partial \sum\nolimits_{m = 1}^M {R_m^{{\rm{appro}}}} }}{{\partial {{\bf{t}}_n}}}$, for each $n = 1,...,N$. Specifically, to obtain $\frac{{\partial \sum\nolimits_{m = 1}^M {R_m^{{\rm{appro}}}} }}{{\partial {{\bf{t}}_n}}}$, we need the results about ${f_{m,1}}(t)$, ${f_{m,2}}(t)$, ${f_{mj,3}}(t)$, $\frac{{\partial {f_{m,1}}(t)}}{{\partial {x_n}}}$ ($\frac{{\partial {f_{m,1}}(t)}}{{\partial {y_n}}}$), $\frac{{\partial {f_{m,2}}(t)}}{{\partial {x_n}}}$ ($\frac{{\partial {f_{m,2}}(t)}}{{\partial {y_n}}}$) and $\frac{{\partial {f_{mj,3}}(t)}}{{\partial {x_n}}}$ ($\frac{{\partial {f_{mj,3}}(t)}}{{\partial {y_n}}}$) for each $m \in {\cal M}$ and $j \ne m$. In particular, for arbitrary $m$, i) computing ${f_{m,1}}(t)$, ${f_{m,2}}(t)$ and ${f_{mj,3}}(t)$ all requires the result of the inverse of the matrix ${{{\overline {\bf{H}} }^H}({\bf{t}})\overline {\bf{H}} ({\bf{t}})}$, the complexity of which is about ${\cal O}\left( {{N^3}} \right)$; ii) computing $\frac{{\partial {f_{m,1}}(t)}}{{\partial {x_n}}}$ ($\frac{{\partial {f_{m,1}}(t)}}{{\partial {y_n}}}$), $\frac{{\partial {f_{m,2}}(t)}}{{\partial {x_n}}}$ ($\frac{{\partial {f_{m,2}}(t)}}{{\partial {y_n}}}$) and $\frac{{\partial {f_{mj,3}}(t)}}{{\partial {x_n}}}$ ($\frac{{\partial {f_{mj,3}}(t)}}{{\partial {y_n}}}$) all requires the result of $\frac{{\partial \overline {{\bf{HH}}} ({\bf{t}})}}{{\partial {x_n}}}$ ($\frac{{\partial \overline {{\bf{HH}}} ({\bf{t}})}}{{\partial {y_n}}}$), the complexity of which is about ${{\cal O}}\left( {{N^2}{M^2} + N{M^3}} \right)$. Therefore, the complexity of computing $\frac{{\partial \sum\nolimits_{m = 1}^M {R_m^{{\rm{appro}}}} }}{{\partial {{\bf{t}}_n}}}$ is about ${\cal O}\left( {{N^3} + {N^2}{M^2} + N{M^3}} \right)$. According to the above analysis, in the $k + 1$-th iteration, the complexity of computing ${\nabla _{{{\bf{t}}^{(k)}}}}\sum\nolimits_{m = 1}^M {R_m^{{\rm{appro}}}}$ is about ${\cal O}\left( {{N^3} + N\left( {{N^2}{M^2} + N{M^3}} \right)} \right)$. Therefore, considering that the PGA method is started from $I$ random antenna positions, the total complexity of Algorithm 1 is about ${\cal{O}}\left( {{K_{{\rm{Iteration}}}}I\left( {{N^3} + N\left( {{N^2}{M^2} + N{M^3}} \right)} \right)} \right)$, where ${{K_{{\rm{Iteration}}}}}$ is the number of iterations for satisfying the convergence condition.

\section{Simulation Results}
This section provides numerical results to verify the effectiveness of our proposed scheme. In the simulations, we consider that there are $M = 4$ users, which are distributed on a circle with a radius of $r = 10^3$ m centered on the BS, and the large-scale fading coefficient can be computed as ${\beta _m} \buildrel \Delta \over = \beta  = {\beta _0}{r^{ - {\alpha _0}}}$, $\forall m$, where ${\beta _0} = {10^{ - 3}}$ is the reference average channel power gain at 1 m and ${{\alpha _0}} = 2$ is the path loss exponent. The transmit power budget of the BS for the $m$-th user is ${P_m} = 10$ dBm, $\forall m$. The Rician-K factor $K_m$ is set as ${K_m} = K$, $\forall m$, providing consistent LoS conditions across users. The elevation and azimuth AoDs of these four users are set as $\left[ {\begin{array}{*{20}{c}}
{{\rm{0}}{\rm{.8676}}}&{{\rm{0}}{\rm{.9879}}}&{{\rm{1}}{\rm{.2720}}}&{{\rm{0}}{\rm{.4021}}}
\end{array}} \right]$ and $\left[ {\begin{array}{*{20}{c}}
{{\rm{0}}{\rm{.2852}}}&{{\rm{1}}{\rm{.1165}}}&{{\rm{1}}{\rm{.0048}}}&{{\rm{1}}{\rm{.2045}}}
\end{array}} \right]$, respectively. In addition, each antenna at the BS can flexibly move in a square with the side length as $L$ (unit: $\lambda $), with $x_n^{\max } - x_n^{\min } = L$ and $y_n^{\max } - y_n^{\min } = L$, $\forall n$. More specifically, we set $x_n^{\min } = (n - 1)(L + 0.5\lambda )$ and $x_n^{\max } = (n - 1)(L + 0.5\lambda ) + L$, where $0.5\lambda $ is the distance between any two adjacent moving regions for avoiding the coupling effect. Also, $y_n^{\min } = 0$ and $y_n^{\max } = L$, $\forall n$. The noise power is fixed as ${\sigma ^2} =  - 90$ dBm. Other simulation-specific parameters will be reflected in the corresponding figure.

\begin{figure}
\centering
\includegraphics[width=8.4cm]{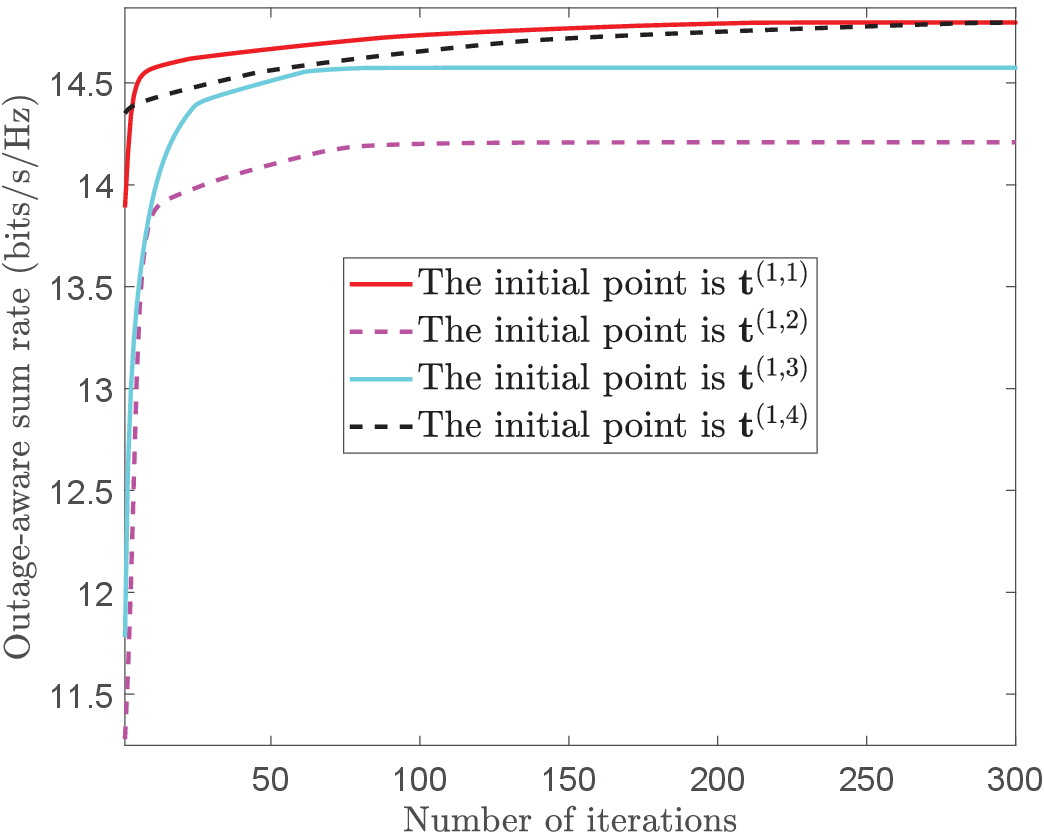}
\captionsetup{font=small}
\caption{The convergence behavior of our proposed PGA method, where $N = 5$, $L = \lambda $, $K = 15$ and $\delta  = 0.2$.} \label{fig:Fig2}
\end{figure}

We first demonstrate the convergence behavior of our proposed PGA method in Fig. 4, where we set the step size as $\varsigma  = 0.015$ and randomly generate $I = 4$ initial antenna positions, i.e.,
\begin{equation} \nonumber
\begin{split}{}
{{\bf{t}}^{(1,1)}} =& \left[ {\begin{array}{*{20}{c}}
{0.4}&{2.3}&{4.7}&{5.5}&{7.4}\\
{0.4}&{0.3}&{0.6}&{0.8}&{0.8}
\end{array}} \right]\lambda, \\
{{\bf{t}}^{(1,2)}} =& \left[ {\begin{array}{*{20}{c}}
{0.8}&{2.5}&{4.2}&{5.8}&{7.1}\\
{0.4}&{0.3}&{0.6}&{0.5}&{0.5}
\end{array}} \right]\lambda, \\
{{\bf{t}}^{(1,3)}} =& \left[ {\begin{array}{*{20}{c}}
{0.7}&{2.1}&{3.8}&{5.5}&{7.15}\\
{0.6}&{0.3}&{0.4}&{0.5}&{0.9}
\end{array}} \right]\lambda, \\
{{\bf{t}}^{(1,4)}} =& \left[ {\begin{array}{*{20}{c}}
{0.5}&{2.5}&{4.5}&{5.22}&{7.13}\\
{0.47}&{0.33}&{0.69}&{0.88}&{0.82}
\end{array}} \right]\lambda.
\end{split}
\end{equation}
From Fig. 4 we can observe that: {\textbf{i)}} For different initial antenna positions, the objective value of (P2) consistently converges to a stationary point after approximately 300 iterations, demonstrating the strong convergence performance of Algorithm 1; {\textbf{ii)}} Compared to the initial points ${{\bf{t}}^{(1,2)}}$ and ${{\bf{t}}^{(1,3)}}$, the algorithm converges to a higher objective value when starting from ${{\bf{t}}^{(1,1)}}$ and ${{\bf{t}}^{(1,4)}}$. As a result, this higher value is selected as the final output of Algorithm 1. In general, as $I$ increases, the performance of Algorithm 1 improves gradually until it eventually saturates, while the computational complexity increases linearly. To balance performance and complexity, we set $I = 5$ in the subsequent simulations.

For comprehensive performance comparisons, we now consider four schemes as benchmarks:
\begin{itemize}
\item[$\bullet$]  \textbf{Fixed-Position Antenna (FPA)}: The BS is equipped with FPA-based uniform linear arrays (ULAs) with $N$ antenna, spaced by $0.5\lambda $. The outage-aware sum rate can be computed immediately with such fixed position deployments.

\item[$\bullet$]  \textbf{Random Antenna Position (RAP)}: Randomly generate ${\bf{t}}$ satisfying (40b) for $10^2$ independent realizations. Obtain the outage-aware sum rate in (40a) for each realization, and then select the maximum value.

 \item[$\bullet$] \textbf{Antenna Selection (AS)}: The BS is equipped with FPA-based ULAs with $2N$ antennas, spaced by $0.5\lambda $. Then, the optimal $N$ antennas are selected via exhaustive search to maximize the outage-aware sum rate in (40a).

 \item[$\bullet$]  \textbf{Rotatable Uniform Linear Array (RULA)}: The BS is equipped with RULAs consisting of $N$ antennas, spaced by $0.5\lambda $. The rotation of the RULAs can be quantized into $10^2$ discrete angles in the range of $\left[ { - \pi ,\pi } \right]$. Then, the optimal rotated angle is selected from $10^2$ realizations to achieve the largest outage-aware sum rate.
\end{itemize}

\begin{figure}
\centering
\includegraphics[width=8cm]{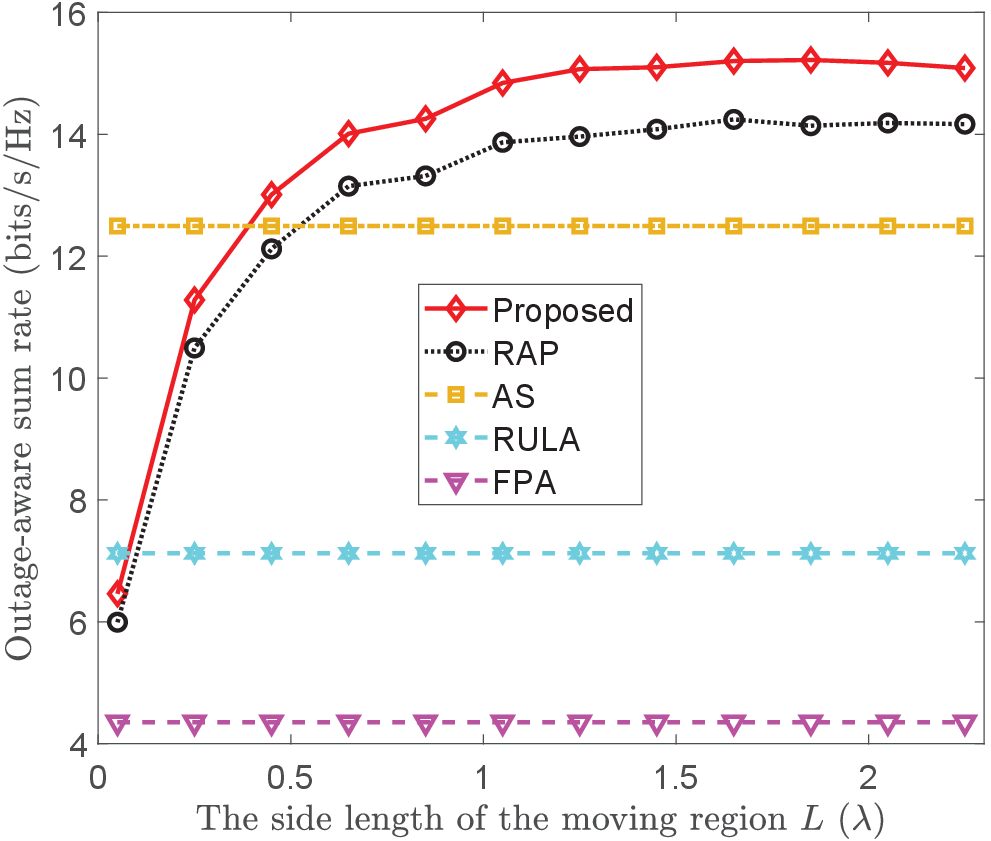}
\captionsetup{font=small}
\caption{Outage-aware sum rate versus the side length of the moving region $L$, where $N = 5$, $K = 15$ and $\delta  = 0.2$.} \label{fig:Fig2}
\end{figure}

Fig. 5 presents the outage-aware sum rate versus the side length $L$ of the moving region of each antenna at the BS, from which we can observe that: \textbf{i)} As $L$ increases, e.g., from $0.05\lambda $ to $1.6\lambda $, each antenna at the BS can be deployed in a larger region to explore the more favorable channel conditions for decreasing the correlation among different multiuser channels. Therefore, the outage-aware sum rate achieved by our proposed scheme and RAP increases with respect to $L$. While as $L$ continues to increase, e.g., from $1.6\lambda $ to $2.25\lambda $, the performance of our proposed scheme and RAP converges to a steady state. This convergence occurs because each element of the statistical downlink channel vector is a function of the sine and cosine of the antenna position. Due to the periodic nature of these trigonometric functions, increasing $L$ beyond a certain point yields saturated returns. Consequently, a finite value of $L$ is sufficient to achieve satisfactory performance. As a comparison, since antenna positions are fixed in the AS, RULA and FPA schemes, the performance of which is not affected by the changeable $L$; \textbf{ii)} When $L$ is small, e.g., from $0.05\lambda $ to $0.4\lambda $, the AS scheme exhibits a performance advantage over our proposed scheme. This is mainly due to that the performance of our scheme is bounded by the inherently low spatial DoF in a constrained region. The AS scheme, however, capitalizes on the hardware redundancy of multiple antenna units to dynamically construct favorable channels through optimal selection, thus providing greater flexibility. While the situation turns into the opposite as $L$ continues to increase. This phenomenon suggests that significant gains can be achieved through a modest expansion of the moving area, avoiding the requirement for deploying costly additional antennas or RF units; \textbf{iii)} The RULA scheme surpasses our proposed scheme only when $L$ is small, e.g., $0.05\lambda $ in Fig. 5. The key reason is that when the antenna arrays, as a rigid body, rotates in the RULA scheme, all AoDs change synchronously by the same amount. This will result in only a limited enhancement in channel reconstruction; \textbf{iv)} For the RAP scheme, the so-called ``best'' antenna positions are selected from multiple random realizations and lacks systematic optimization. This inherently heuristic approach leads to suboptimal selections, resulting in a rate performance inferior to our proposed scheme; \textbf{v)} The FPA scheme, with the fixed antenna positions at the BS, offers no flexibility to exploit additional spatial DoF, which fundamentally limits it to the worst performance.

\begin{figure}
\centering
\includegraphics[width=8cm]{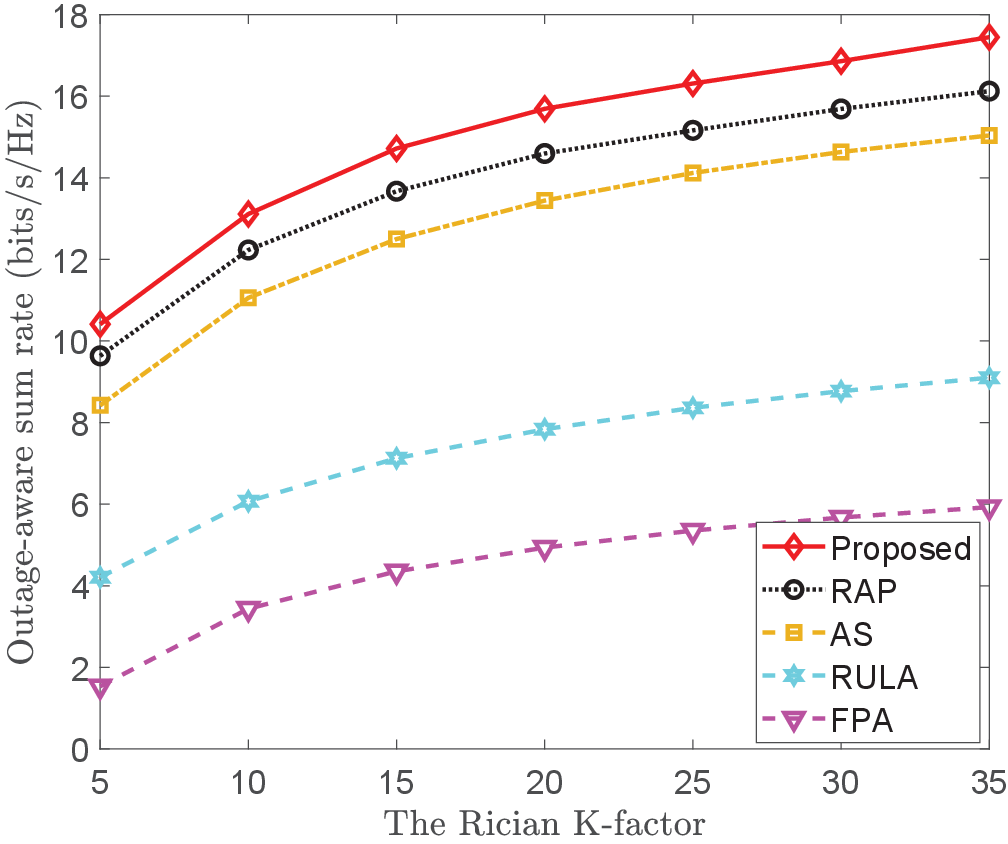}
\captionsetup{font=small}
\caption{Outage-aware sum rate versus the Rician K-factor, where $N = 5$, $L = \lambda $ and $\delta  = 0.2$.} \label{fig:Fig2}
\end{figure}

Fig. 6 shows the outage-aware sum rate versus the Rician K-factor, from which we can observe that: \textbf{i)} As $K$ increases, the LoS component becomes dominant in the downlink channel vector, while the NLoS component diminishes. As a result, the mutual interference caused by the NLoS path at each user decreases, whereas the useful average received power at each user increases, which will lead to sustained performance improvement with respect to $K$ for all schemes. In particular, as $K$ tends to infinity, there is no random NLoS component in the downlink channel vector, indicating that no outage event will occur. Therefore, based on (8), given ${\bf{t}}$, the non-outage-aware sum rate will become $\sum\nolimits_{m = 1}^M {{{\log }_2}\left( {1 + {P_m}{{\left| {{{\overline {\bf{h}} }_m}({\bf{t}}){\bf{w}}_m^{{\rm{ZF}}}({\bf{t}})} \right|}^2}/{\sigma ^2}} \right)} $; \textbf{ii)} The sum rate performance of our proposed scheme is approximately three times that of traditional FPA scheme, implying potential advantages achieved by reconfiguring channels via strategic antenna movements.

\begin{figure}
\centering
\includegraphics[width=8cm]{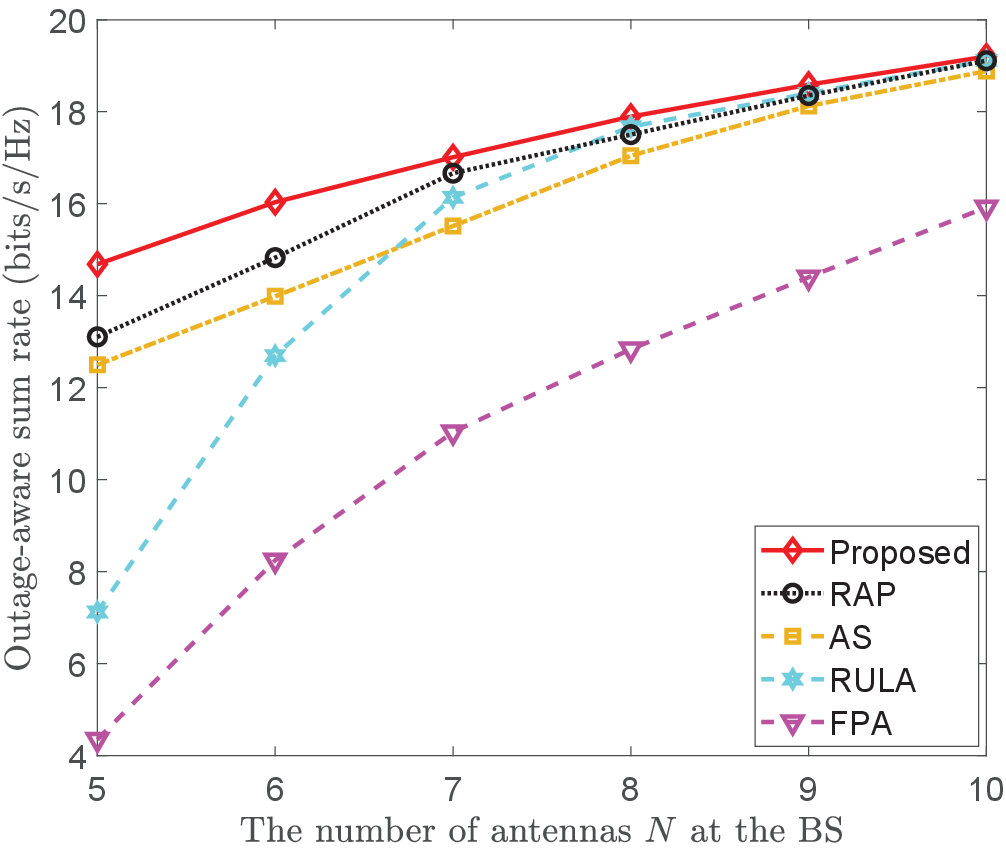}
\captionsetup{font=small}
\caption{Outage-aware sum rate versus the number of antennas ($N$) at the BS, where $L = \lambda $, $K = 15$ and $\delta  = 0.2$.} \label{fig:Fig2}
\end{figure}

\begin{figure}
\centering
\includegraphics[width=8cm]{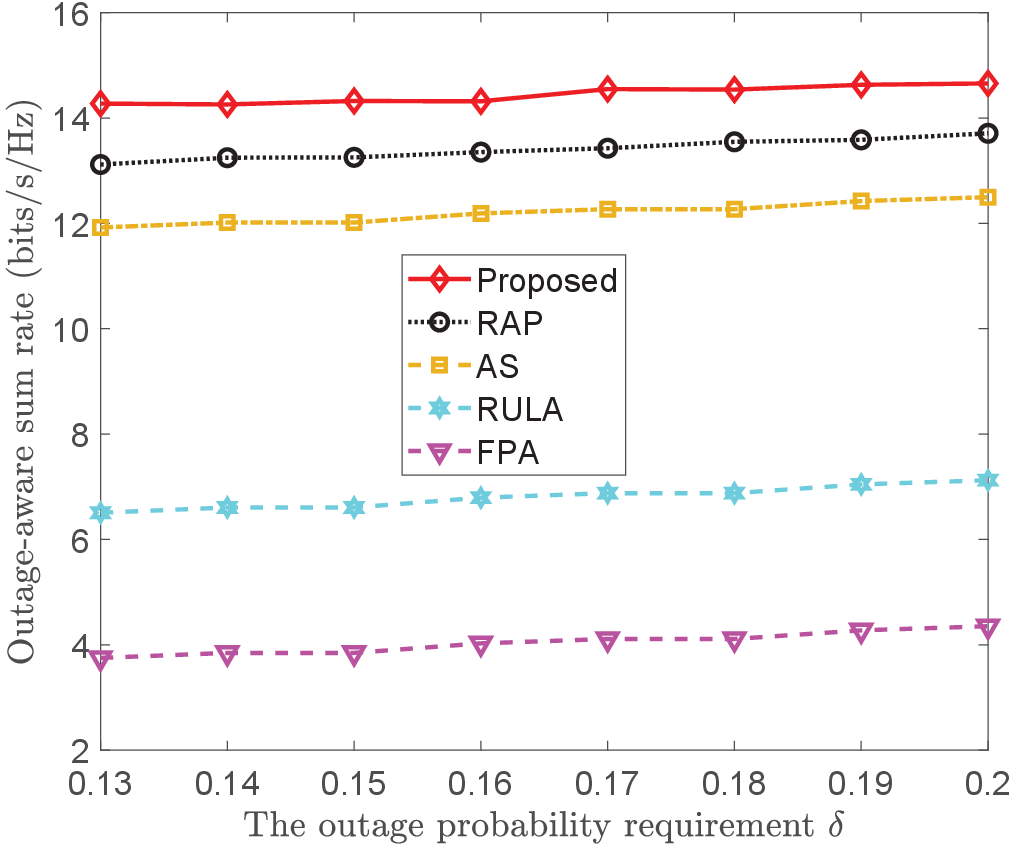}
\captionsetup{font=small}
\caption{Outage-aware sum rate versus the outage probability requirement $\delta $, where $N = 5$, $L = \lambda $ and $K = 15$.} \label{fig:Fig2}
\end{figure}

Fig. 7 provides the outage-aware sum rate versus the number of antennas $N$ at the BS, from which we can observe that: \textbf{i)} As $N$ increases, the BS can employ enhanced beamforming to boost the average received power at each user, thereby achieving a higher spatial diversity. Hence, obviously the sum rate performance of all scheme improves with respect to $N$; \textbf{ii)} The performance gap between our proposed scheme and the benchmarks (AS, RULA, FPA) narrows noticeably with the growth of $N$. This phenomenon occurs because a larger $N$ has inherently reduced the correlation among the statistical LoS components across different channels. Consequently, compared to rotating antenna arrays in RULA or just fixing antenna positions in FPA, the relative benefit of antenna movements in decorrelating channels diminishes, thereby lessening the advantage of our proposed scheme.

Fig. 8 shows the outage-aware sum rate versus the outage probability requirement $\delta $ at users, from which we can observe that: \textbf{i)} As $\delta $ increases, the BS can accordingly enhance its transmission rate to accommodate the greater tolerance for outage events. Therefore, the outage-aware sum rate of all schemes increases with respect to $\delta $; \textbf{ii)} As the outage probability requirement becomes more stringent, our proposed scheme maintains a stable sum rate, demonstrating significantly higher robustness than conventional FPA. This result validates the effectiveness of jointly optimizing antenna positions and transmit beamforming to enhance system robustness.

\section{Conclusions}
In this paper, we investigated the outage-aware sum rate maximization problem for MAs-enabled MISO systems, operating under the practical delay-sensitive scenario where only statistical CSI is available at the BS. To tackle the highly non-convex optimization challenge stemming from the intractable CDF of the received SINR, we developed a comprehensive solution framework. The key aspects to our approach were the exploitation of the Taylor expansion method for deriving the tight mean and variance for the SINR and the subsequent utilization of the Laguerre series to obtain a tight CDF expression. Based on this analytical foundation, we then obtained a closed-form outage-aware sum rate expression, and further proposed an efficient PGA-based algorithm to optimize antenna positions for performance maximization. Extensive numerical results have validated the superiority of our proposed scheme over conventional FPA scheme and other competitive benchmarks.


\normalem
\bibliographystyle{IEEEtran}
\bibliography{IEEEabrv,mybib}

\begin{thebibliography}{10}
\providecommand{\url}[1]{#1}
\csname url@samestyle\endcsname
\providecommand{\newblock}{\relax}
\providecommand{\bibinfo}[2]{#2}
\providecommand{\BIBentrySTDinterwordspacing}{\spaceskip=0pt\relax}
\providecommand{\BIBentryALTinterwordstretchfactor}{4}
\providecommand{\BIBentryALTinterwordspacing}{\spaceskip=\fontdimen2\font plus
\BIBentryALTinterwordstretchfactor\fontdimen3\font minus
  \fontdimen4\font\relax}
\providecommand{\BIBforeignlanguage}[2]{{%
\expandafter\ifx\csname l@#1\endcsname\relax
\typeout{** WARNING: IEEEtran.bst: No hyphenation pattern has been}%
\typeout{** loaded for the language `#1'. Using the pattern for}%
\typeout{** the default language instead.}%
\else
\language=\csname l@#1\endcsname
\fi
#2}}
\providecommand{\BIBdecl}{\relax}
\BIBdecl

\bibitem{An_Overview}
A.~PAULRAJ, D.~GORE, R.~NABAR, and H.~BOLCSKEI, ``An overview of {MIMO}
  communications - a key to gigabit wireless,'' \emph{Proc. IEEE}, vol.~92,
  no.~2, pp. 198--218, 2004.

\bibitem{Shift_MIMO}
D.~Gesbert, M.~Kountouris, R.~W. Heath, C.-b. Chae, and T.~Salzer, ``Shifting
  the {MIMO} paradigm,'' \emph{IEEE Signal Process. Mag.}, vol.~24, no.~5, pp.
  36--46, 2007.

\bibitem{Massivea}
E.~G. Larsson, O.~Edfors, F.~Tufvesson, and T.~L. Marzetta, ``Massive {MIMO}
  for next generation wireless systems,'' \emph{IEEE Commun. Mag.}, vol.~52,
  no.~2, pp. 186--195, 2014.

\bibitem{Massivet}
Y.~Zeng and R.~Zhang, ``Millimeter wave {MIMO} with lens antenna array: A new
  path division multiplexing paradigm,'' \emph{IEEE Trans. Commun.}, vol.~64,
  no.~4, pp. 1557--1571, 2016.

\bibitem{Guojie1}
G.~Hu, Q.~Wu, D.~Xu, K.~Xu, J.~Si, Y.~Cai, and N.~Al-Dhahir, ``Movable
  antennas-assisted secure transmission without eavesdroppers' instantaneous
  {CSI},'' \emph{IEEE Trans. Mobile Comput.}, vol.~23, no.~12, pp.
  14\,263--14\,279, 2024.

\bibitem{Xiqi_Gao}
C.~Sun, X.~Gao, S.~Jin, M.~Matthaiou, Z.~Ding, and C.~Xiao, ``Beam division
  multiple access transmission for massive {MIMO} communications,'' \emph{IEEE
  Trans. Commun.}, vol.~63, no.~6, pp. 2170--2184, 2015.

\bibitem{Massive2}
K.~Dovelos, M.~Matthaiou, H.~Q. Ngo, and B.~Bellalta, ``Channel estimation and
  hybrid combining for wideband terahertz massive {MIMO} systems,'' \emph{IEEE
  J. Sel. Areas Commun.}, vol.~39, no.~6, pp. 1604--1620, 2021.

\bibitem{DLL1}
M.~Cui and L.~Dai, ``Channel estimation for extremely large-scale {MIMO}:
  Far-field or near-field?'' \emph{IEEE Trans. Commun.}, vol.~70, no.~4, pp.
  2663--2677, 2022.

\bibitem{DLL2}
Z.~Wu and L.~Dai, ``Multiple access for near-field communications: {SDMA} or
  {LDMA}?'' \emph{IEEE J. Sel. Areas Commun.}, vol.~41, no.~6, pp. 1918--1935,
  2023.

\bibitem{ZLP_survey}
L.~Zhu, W.~Ma, W.~Mei, Y.~Zeng, Q.~Wu, B.~Ning, Z.~Xiao, X.~Shao, J.~Zhang, and
  R.~Zhang, ``A tutorial on movable antennas for wireless networks,''
  \emph{IEEE Commun. Surv. Tuts.}, pp. 1--1, 2025.

\bibitem{Qingqing_arxiv}
Q.~Wu, Z.~Zheng, G.~Ying, W.~Mei, X.~Wei, and B.~Ning, ``Integrating movable
  antennas and intelligent reflecting surfaces ({MA-IRS}): Fundamentals,
  practical solutions, and opportunities,'' 2025, [Online] Available:
  \url{https://arxiv.org/abs/2506.14636}.

\bibitem{Z_WCM}
J.~Zheng, J.~Zhang, H.~Du, D.~Niyato, S.~Sun, B.~Ai, and K.~B. Letaief,
  ``Flexible-position {MIMO} for wireless communications: Fundamentals,
  challenges, and future directions,'' \emph{IEEE Wireless Commun.}, vol.~31,
  no.~5, pp. 18--26, 2024.

\bibitem{ZLP_MODEL}
L.~Zhu, W.~Ma, and R.~Zhang, ``Modeling and performance analysis for movable
  antenna enabled wireless communications,'' \emph{IEEE Tran. Wireless
  Commun.}, vol.~23, no.~6, pp. 6234--6250, 2024.

\bibitem{MWYY_TWC}
W.~Ma, L.~Zhu, and R.~Zhang, ``M{IMO} capacity characterization for movable
  antenna systems,'' \emph{IEEE Tran. Wireless Commun.}, vol.~23, no.~4, pp.
  3392--3407, 2024.

\bibitem{gao2025integrating}
Y.~Gao, Q.~Wu, W.~Mei, G.~Chen, W.~Chen, and Z.~Zheng, ``Integrating movable
  antennas and intelligent reflecting surfaces for coverage enhancement,''
  2025, [Online] Available: \url{https://arxiv.org/abs/2506.21375}.

\bibitem{Wanghaohao}
H.~Wang, Q.~Wu, and W.~Chen, ``Movable antenna enabled interference network:
  Joint antenna position and beamforming design,'' \emph{IEEE Wireless Commun.
  Lett.}, vol.~13, no.~9, pp. 2517--2521, 2024.

\bibitem{ZLP23}
L.~Zhu, W.~Ma, B.~Ning, and R.~Zhang, ``Movable-antenna enhanced multiuser
  communication via antenna position optimization,'' \emph{IEEE Trans. Wireless
  Commun.}, vol.~23, no.~7, pp. 7214--7229, 2024.

\bibitem{GUOJIE_CL}
G.~Hu, Q.~Wu, K.~Xu, J.~Ouyang, J.~Si, Y.~Cai, and N.~Al-Dhahir, ``Fluid
  antennas-enabled multiuser uplink: A low-complexity gradient descent for
  total transmit power minimization,'' \emph{IEEE Commun. Lett.}, vol.~28,
  no.~3, pp. 602--606, 2024.

\bibitem{XZY1}
Z.~Xiao, X.~Pi, L.~Zhu, X.-G. Xia, and R.~Zhang, ``Multiuser communications
  with movable-antenna base station: Joint antenna positioning, receive
  combining, and power control,'' \emph{IEEE Trans. Wireless Commun.}, vol.~23,
  no.~12, pp. 19\,744--19\,759, 2024.

\bibitem{GUOJIE_TWOTIMESCALE}
G.~Hu, Q.~Wu, G.~Li, D.~Xu, K.~Xu, J.~Si, Y.~Cai, and N.~Al-Dhahir,
  ``Two-timescale design for movable antenna array-enabled multiuser uplink
  communications,'' \emph{IEEE Trans. Veh. Technol.}, vol.~74, no.~3, pp.
  5152--5157, 2025.

\bibitem{XUHAO_TCOM}
H.~Xu, K.-K. Wong, W.~K. New, F.~R. Ghadi, G.~Zhou, R.~Murch, C.-B. Chae,
  Y.~Zhu, and S.~Jin, ``Capacity maximization for {FAS}-assisted multiple
  access channels,'' \emph{IEEE Trans. Commun.}, vol.~73, no.~7, pp.
  4713--4731, 2025.

\bibitem{XIAODAN1}
X.~Shao, Q.~Jiang, and R.~Zhang, ``6{D} movable antenna based on user
  distribution: Modeling and optimization,'' \emph{IEEE Trans. Wireless
  Commun.}, vol.~24, no.~1, pp. 355--370, 2025.

\bibitem{XIAODAN2}
X.~Shao, R.~Zhang, Q.~Jiang, and R.~Schober, ``6{D} movable antenna enhanced
  wireless network via discrete position and rotation optimization,''
  \emph{IEEE J. Sel. Areas Commun.}, vol.~43, no.~3, pp. 674--687, 2025.

\bibitem{WUYIFEI_TCOM}
Y.~Wu, D.~Xu, D.~W.~K. Ng, W.~Gerstacker, and R.~Schober, ``Globally optimal
  movable antenna-enabled multiuser communication: Discrete antenna
  positioning, power consumption, and imperfect {CSI},'' \emph{IEEE Trans.
  Commun.}, pp. 1--1, 2025.

\bibitem{IOT}
J.~Guo, S.~Yang, X.~Dong, J.~Yang, J.~Deng, Z.~Zhang, and C.~Yuen, ``Flexible
  cylindrical arrays with movable antennas for {MISO} system: Beamforming and
  position optimization,'' \emph{IEEE Internet Things J.}, vol.~12, no.~17, pp.
  35\,394--35\,405, 2025.

\bibitem{MA_NEAR1}
X.~Pi, L.~Zhu, H.~Mao, Z.~Xiao, X.-G. Xia, and R.~Zhang, ``Movable antenna
  enabled near-field {MU}-{MIMO} communication,'' \emph{IEEE Wireless Commun.
  Lett.}, pp. 1--1, 2025.

\bibitem{MA_NEAR2}
J.~Ding, L.~Zhu, Z.~Zhou, B.~Jiao, and R.~Zhang, ``Near-field multiuser
  communications aided by movable antennas,'' \emph{IEEE Wireless Commun.
  Lett.}, vol.~14, no.~1, pp. 138--142, 2025.

\bibitem{ZHENZY}
Z.~Zheng, Q.~Wu, W.~Chen, and G.~Hu, ``Two-timescale design for movable
  antennas enabled-multiuser {MIMO} systems,'' \emph{IEEE Trans. Commun.}, pp.
  1--1, 2025.

\bibitem{CHENXT}
X.~Chen, B.~Feng, Y.~Wu, D.~W.~K. Ng, and R.~Schober, ``Two-timescale sum-rate
  maximization for movable antenna enhanced systems,'' \emph{IEEE Trans.
  Wireless Commun.}, pp. 1--1, 2025.

\bibitem{Delay_sensitive}
R.~Han, Y.~Yu, L.~Bai, J.~Wang, J.~Choi, and W.~Zhang, ``Effective capacity
  analysis of delay-sensitive communications in {NOMA} systems,'' \emph{IEEE
  Trans. Wireless Commun.}, vol.~23, no.~4, pp. 3665--3675, 2024.

\bibitem{Outage1}
Y.~Lu, K.~Xiong, P.~Fan, B.~Ai, and Z.~Zhong, ``Outage-constrained sum
  transmission rate maximization in {RIS}-assisted {MISO} systems,'' \emph{IEEE
  Trans. Wireless Commun.}, vol.~23, no.~4, pp. 2505--2518, 2024.

\bibitem{Outage2}
M.-M. Zhao, A.~Liu, and R.~Zhang, ``Outage-constrained robust beamforming for
  intelligent reflecting surface aided wireless communication,'' \emph{IEEE
  Trans. Signal Process.}, vol.~69, pp. 1301--1316, 2021.

\bibitem{GUOJIE_TCOM}
G.~Hu, Z.~Li, J.~Si, K.~Xu, D.~Xu, Y.~Cai, and N.~Al-Dhahir, ``Maxmin fairness
  for {UAV}-enabled proactive eavesdropping with jamming over distributed
  transmit beamforming-based suspicious communications,'' \emph{IEEE Trans.
  Commun.}, vol.~71, no.~3, pp. 1595--1614, 2023.

\bibitem{HUAMENG}
M.~Hua, L.~Yang, Q.~Wu, and A.~L. Swindlehurst, ``3{D} {UAV} trajectory and
  communication design for simultaneous uplink and downlink transmission,''
  \emph{IEEE Trans. Commun.}, vol.~68, no.~9, pp. 5908--5923, 2020.

\end{thebibliography}

\end{document}